\documentclass{iopjournal}

\usepackage{ragged2e}
\usepackage{amssymb}
\usepackage{amsmath}
\usepackage{pgfplots}

\newcommand\bb{\boldsymbol{b}}
\newcommand\bB{\boldsymbol{B}}

\newcommand{\sep}{,\ }
\pgfplotsset{compat=1.18}

\begin{document}
\justifying
\articletype{Paper} %	 e.g. Paper, Letter, Topical Review...

\title{A High-order piecewise field-aligned triangular finite element method for electromagnetic gyrokinetic particle  simulations of tokamak plasmas with open field lines}

\author{Zhixin Lu$^{1*}$, Guo Meng$^{1*}$, Eric Sonnendrücker$^1$, Roman Hatzky$^1$, Giorgio Daneri$^1$,  Gengxian Li$^1$, Peiyou Jiang$^1$, Klaus Reuter$^2$, Matthias Hoelzl$^1$}

\affil{$^1${Max Planck Institute for Plasma Physics, Boltzmannstr. 2,  Garching, 85748, Germany}}
\affil{$^2${Max Planck Computing and Data Facility, Giessenbachstrasse 2,  Garching, 85748, Germany}}

\affil{$^*$Author to whom any correspondence should be addressed (zhixin.lu@ipp.mpg.de; guo.meng@ipp.mpg.de).}

%\email{name@institution.org}

\keywords{Whole volume simulations\sep Electromagnetic gyrokinetic simulation\sep particle-in-cell\sep turbulence\sep Alfv\'en modes\sep finite element method\sep triangular meshes\sep field-aligned\sep tokamak\sep magnetic confinement fusion\sep divertor tokamak\sep open field lines}

\begin{abstract}
\justifying 
A high-order piecewise field-aligned triangular finite element method is developed and implemented for global electromagnetic gyrokinetic particle-in-cell simulations of tokamak plasmas with open field lines. The approach combines locally field-aligned finite element basis functions with unstructured $C^{1}$ triangular meshes in cylindrical coordinates, enabling whole-volume simulations with substantially reduced computational effort, while avoiding the grid distortion associated with globally field-aligned coordinates and the associated singularity at the separatrix of diverted plasmas. The formulation is compatible with both $\delta f$ and full-$f$ models and employs  mixed-variable representations, along with a generalized pullback scheme, to control numerical cancellation in electromagnetic simulations. The method is implemented in the TRIMEG-C1 code and demonstrated using linear and nonlinear electromagnetic simulations of the TCV-X21 configuration. The results indicate that the approach accurately captures the key features of electromagnetic ion-temperature-gradient and kinetic ballooning mode physics, including the separatrix regions in the simulation, thereby providing a robust framework for whole-volume electromagnetic gyrokinetic simulations in realistic tokamak geometries.
\end{abstract}

\section{Introduction}
\label{sec:intro}

Electromagnetic effects are essential for accurately describing microinstabilities and turbulence in magnetically confined fusion plasmas, including ion-temperature-gradient (ITG) modes, kinetic ballooning modes (KBMs), and Alfv\'enic fluctuations. Their importance increases at finite plasma pressure $\beta$, particularly in global simulations that span both core and edge regions. First-principles electromagnetic gyrokinetic simulations are therefore a key tool for predictive modelling of tokamak plasmas \cite{mishchenko2023numerical,lanti2020orb5,taimourzadeh2019verification}.
Global gyrokinetic particle-in-cell (PIC) simulations in realistic geometry remain numerically challenging due to strong scale anisotropy (the wave vector component along the magnetic field line is much smaller than the perpendicular component, $k_\parallel \ll k_\perp$), magnetic shear, and the presence of open magnetic field lines outside the separatrix. While globally field-aligned coordinate systems reduce numerical stiffness along magnetic field lines, they often suffer from severe grid distortion in whole-volume simulations and exhibit a singularity on the separatrix. Finite element methods on unstructured meshes offer greater geometric flexibility \cite{chang2017fast}, especially when combined with high-order basis functions that improve numerical accuracy \cite{lu2024gyrokinetic}.

To exploit field-line alignment while avoiding global grid distortion, a \textit{piecewise} field-aligned finite element method has been proposed, in which basis functions are locally aligned with magnetic field lines within each toroidal subdomain \cite{lu2025piecewise}. It has been originally implemented in the TRIMEG-GKX code for core plasma without open field lines \cite{lu2026cpc}. This approach preserves continuity along field lines and significantly reduces numerical errors associated with parallel dynamics. Similar ideas have been used for Finite Difference schemes. Specifically, the flux-coordinate independent (FCI) approach has previously been developed to improve efficiency and accuracy along the magnetic field \cite{hariri2013flux,stegmeir2018grillix}, and the shifted metric procedure has been developed to reduce the grid distortion \cite{scott2001shifted}. Our scheme is formulated using the finite element method (FEM), thereby inheriting the advantageous features of FEM, such as its ability to handle complex, curved, and irregular geometries and refinement. Additionally, mixed-variable formulations combined with pullback schemes are essential for controlling numerical cancellation in electromagnetic gyrokinetic PIC simulations \cite{mishchenko2014pullback,hatzky2019reduction,lu2025generalized}.

In this work, we extend the piecewise field-aligned finite element method (PFAFEM) to global electromagnetic gyrokinetic simulations with \textit{open} field lines. The PFAFEM enables us to use an identical mesh structure at different toroidal locations, which differs from that in XGC \cite{moritaka2019development}. The formulation is implemented in cylindrical coordinates in the TRIMEG-C1 code using high-order $C^1$ finite elements on triangular meshes in the poloidal cross section. It is compatible with both $\delta f$ and full-$f$ models. The method incorporates generalized pullback schemes and an efficient iterative treatment of Amp\`ere’s law. Its capabilities are demonstrated through electromagnetic simulations of the TCV configuration, which capture key features of ITG and KBM physics in both the core and near-separatrix regions. 
The remainder of the paper is organized as follows. Section~\ref{sec:model} introduces the gyrokinetic model and electromagnetic field equations. Section~\ref{sec:numeric_scheme} describes the piecewise field-aligned finite element formulation and the corresponding field solvers. Simulation setup and numerical results are presented in Section~\ref{sec:results}, followed by conclusions in Section~\ref{sec:conclusion}.

\section{Models and equations}
\label{sec:model}
The mixed variable–pullback scheme has been studied extensively in previous work \cite{mishchenko2014pullback,hatzky2019reduction,lanti2020orb5} and was recently generalized to allow for various choices of the evolution equation for the symplectic part of $\delta A_\parallel$ \cite{lu2025generalized}. Two such schemes that are most suitable for TRIMEG-C1 have been implemented in this work, as introduced in the following.

\subsection{Physics equations using mixed variables}
Using the mixed variable scheme, the parallel component $\delta A_\|$ of the perturbed magnetic potential is decomposed into a symplectic part and a Hamiltonian part \cite{mishchenko2014pullback},
\begin{equation}
\label{eq:AsAh}
    \delta A_\| =\delta A_\|^{\rm{s}} + \delta A_\|^{\rm{h}} \;\;.
\end{equation}
The shifted parallel velocity coordinate of the gyrocenter $u_\|$ is defined as 
\begin{equation}
    u_\|=v_\|+\frac{q_s}{m_s}\langle\delta A_\|^{\rm{h}}\rangle\;\;,
\end{equation}
where $v_\|$ is the parallel velocity, $q_s$ and $m_s$ are the charge and mass of species $s$, respectively, the subscript ``$s$'' represents the different particle species, and $\langle\ldots\rangle$ indicates the gyro average.

The gyrocenter equations of motion are consistent with previous work \cite{mishchenko2014pullback,hatzky2019reduction,lanti2020orb5,mishchenko2023global,kleiber2024euterpe}, {
\begin{eqnarray}
\label{eq:dRdt}
 \dot{\boldsymbol R}_0 
  &=& u_\| {\boldsymbol b}^*_0 + \frac{m\mu}{qB^*_\|} {\boldsymbol b}\times\nabla B \;\;, 
  \\
  \dot u_{\|,0}
  &=& -\mu {\boldsymbol b}^*_0\cdot \nabla B \;\;,
  \\
  \delta\dot{\boldsymbol R}
  &=& \frac{{\boldsymbol b}}{B^*_\|}\times \nabla \langle \delta\Phi -u_\| \delta A_\|\rangle 
  -\frac{q_s}{m_s}\langle\delta A^{\rm h}_\|\rangle {\boldsymbol b}^*\;\;, 
  \\
  \label{eq:dudt}
  \delta \dot u_\|
  &=&  -\frac{q_s}{m_s} \left({\boldsymbol b}^*\cdot\nabla\langle\delta\Phi-u_\|\delta A^{\rm{h}}_\|\rangle +\partial_t\langle\delta A_\|^{\rm{s}}\rangle \right) \nonumber\\
  &&-\frac{\mu}{B^*_\|}{\boldsymbol b}\times\nabla B\cdot\nabla\langle\delta A_\|^{\rm{s}}\rangle \;\;,  
\end{eqnarray}
where the magnetic moment $\mu=v_\perp^2/(2B)$, 
${\boldsymbol b}^*={\boldsymbol b}_0^*+\nabla\times({\boldsymbol b}\langle\delta A_\|^{\rm s}\rangle)/B_\|^*
\approx{\boldsymbol b}_0^*+\nabla\langle\delta A_\|^{\rm s}\rangle\times{\boldsymbol b}/B_\|^*$, ${\boldsymbol b}^*_0={\boldsymbol b}+(m_s/q_s)u_\|\nabla\times{\boldsymbol b}/B_\|^*$, ${\boldsymbol b}={\boldsymbol B}/B$, $\boldsymbol{B}$ is the equilibrium magnetic field, $B_\|^*=B+(m_s/q_s)u_\|{\boldsymbol b}\cdot(\nabla\times{\boldsymbol b})$. 
In obtaining the approximated form of ${\bb}^*$, we made use of the following simplification of the  perturbed magnetic field $
    \bB_1 =\nabla\times (\delta A_\parallel \bb) 
    = \nabla \delta A_\parallel \times \bb + \delta A_\parallel \nabla \times \bb 
    =\nabla \delta A_\parallel \times \bb + \delta A_\parallel (\bb \times \boldsymbol{\kappa} + (\bb\cdot \nabla \times \bb) \bb) \nonumber 
    \approx \nabla \delta A_\parallel \times \bb$
where $\boldsymbol{\kappa}=(\bb\cdot\nabla)\bb$ is the curvature of the equilibrium magnetic field lines. The modified magnetic field is then
$\bB^*=\bB+\bB_1+ (m_s/q_s) u_\parallel \nabla \times \bb$.
Note that $\nabla \cdot \bB^* $ is not exactly zero, since $\nabla \cdot \bB_1 \neq 0$ due to the approximation.
The expression of the gyrocenter equation of motion Eqs.~\eqref{eq:dRdt}--\eqref{eq:dudt} in $(R,\phi,Z)$ coordinates used in TRIMEG-C1 is listed in  \cite{lu2024gyrokinetic}.

The distribution function is decomposed into the equilibrium one, and the perturbed one $f=f_0+\delta f$, and $\delta f$ is solved along the particle trajectory using the standard procedure \cite{lanti2020orb5,hatzky2019reduction,mishchenko2014pullback,lu2023full}. In particle simulations, $\delta f$ is represented by the numerical markers
\begin{eqnarray}
\label{eq:deltaf}
    \delta f=\frac{\langle n\rangle_V V_{\rm tot}}{N_{\rm mark}}\sum_{p=1}^{N_{\rm mark}} \frac{w_p}{J_z} \delta({\bf z}-{\bf z}_p) \;\;,
\end{eqnarray}
where $N_{\rm mark}$ is the marker number, $ \langle\ldots\rangle_V$ indicates the volume average, and $V_{\rm tot}$ is the total volume.

The linearized quasi-neutrality equation in the long-wavelength approximation is as follows, 
\begin{equation}
\label{eq:poisson0}
    -\nabla\cdot\left( \sum_s\frac{q_s n_{0s}}{B\omega_{{\rm c}s}} \nabla_\perp\delta\Phi \right) = \sum_s q_s \delta n_{s,v} \;\;,
\end{equation}
where the gyrocenter density $\delta n_{s,v}$ is calculated using $\delta f_s({\boldsymbol R},v_\|,\mu)$ (indicated as $\delta f_{s,v}$), namely, $\delta n_{s,v}({\boldsymbol{x} })=\int {\rm d}^6 z\delta f_{s,v}\delta(\boldsymbol{R}  + \boldsymbol{\rho} - \boldsymbol{x} )$. Here, $\boldsymbol{x} $ and $\boldsymbol{R} $ denote the particle position vector and gyrocenter position vector, respectively, and $\boldsymbol{\rho}$ represents the Larmor radius.
In Eq.~\eqref{eq:poisson0}, $\omega_{{\rm c}s}$ is the cyclotron frequency of species $s$, and in this work, we ignore the perturbed electron polarization density on the left-hand side. 
When the $\delta f$ scheme is adopted, $\delta f_{s,v}$ is obtained from $\delta f_{s,u}$ as follows, with the linear approximation of the pullback scheme,
\begin{eqnarray}
    & \delta f_{s,v} = \delta f_{s,u} +  \frac{q_s\left\langle\delta A^{\rm{h}}_{\|} \right\rangle}{m_s}\frac{\partial f_{0s}}{\partial v_\|}
    \xrightarrow[f_{0s}=f_{\rm{M}}]{\text{Maxwellian}}
     \delta f_{s,u} -  \frac{ v_\|}{T_s}  q_s\left\langle\delta A^{\rm{h}}_{\|} \right\rangle f_{0s}\;\;.
\end{eqnarray}

Amp\`ere's law in $v_\|$ space is 
$-\nabla^2_\perp\delta A_\| = \mu_0 \delta j_{\|,v}$
where $\delta j_{\|,v}({\boldsymbol{x} })=\sum_s q_s \int {\rm d}^6 z\delta f_{s,v}\delta(\boldsymbol{R}  + \boldsymbol{\rho} - \boldsymbol{x} )v_\|$. It is solved in the mixed variable space. 
Using an iterative scheme \cite{chen2007electromagnetic,mishchenko2017mitigation,hatzky2019reduction,lu2023full}, the asymptotic solution is expressed as $\delta A^{\rm{h}}_{\|}=\sum_{I=0}^\infty\delta A^{\rm{h}}_{\|,I}$,
where $|\delta A^{\rm{h}}_{\|,I+1}/\delta A^{\rm{h}}_{\|,I}|\ll1$ is assured by the fact that the analytical skin depth term is close to the exact one. 
Amp\`ere's law is solved iteratively as follows,
\begin{eqnarray}
\label{eq:ampere_h0}
    \left(\nabla^2_\perp-\sum_s\frac{1}{d_{s}^2}\right)\delta A_{\|,0}^{\rm{h}} 
    &=& -\nabla^2_\perp\delta A_{\|}^{\rm{s}} - \mu_0 \delta j_{\|} \;\;, \\
\label{eq:ampere_iterative}
    \left(\nabla^2_\perp-\sum_s\frac{1}{d_{s}^2}\right)\delta A_{\|,I}^{\rm{h}} 
    &=&-\sum_s\frac{1}{d_{s}^2}\delta A^{\rm{h}}_{\|,I-1} 
    + \sum_s\frac{1}{d_{s}^2} \overline{\langle\delta A_{\|,I-1}^{\rm{h}}\rangle}\;\;, \\
    \overline{\langle\delta A_{\|,I-1}^{\rm{h}}\rangle}
    &=&\frac{2}{n_0 v_{\rm{t}s}^2}\int \mathrm{d}z^6 v_\|^2 f_{0s}  \langle\delta A^{\rm{h}}_{\|,I-1} \rangle\delta(\boldsymbol{R}  + \boldsymbol{\rho} - \boldsymbol{x} )  \;\;\text{ for $\delta f$}\;\;, \\
\label{eq:A2ndavg_fullf}
    \overline{\langle\delta A_{\|,I-1}^{\rm{h}}\rangle}
    &=&\frac{1}{n_0}\int \mathrm{d}z^6 f_{s,v}  \langle\delta A^{\rm{h}}_{\|,I-1} \rangle \delta(\boldsymbol{R}  + \boldsymbol{\rho} - \boldsymbol{x} ) \;\;\text{ for full $f$  \;\;,}
\end{eqnarray}
where  $I=1,2,3,\ldots$ and details of the $\delta f$ model and full $f$ model can be found in the previous works \cite{hatzky2019reduction,lu2023full}, $v_{ts}=\sqrt{2T_s/m_s}$, $T_s$ is the temperature, $d_s=c/\omega_{\mathrm{p}s}$ is the skin depth, $\omega_{\mathrm{p}s}=\sqrt{n_sq_s^2/(m_s\varepsilon_0)}$ is the plasma frequency of species "s".

\subsection{Generalized mixed variable-pullback scheme}
The original work of the pullback scheme for the $\delta f$ method can be found in the previous work \cite{mishchenko2014pullback} as follows,
\begin{align}
\label{eq:pullback_A}
    & \delta A^{\rm{s}}_{\|,\rm{new}} = \delta A^{\rm{s}}_{\|,\rm{old}} + \delta A^{\rm{h}}_{\|,\rm{old}}  \;\;, \\
\label{eq:pullback_v}
    & u_{\|,\rm{new}} = u_{\|,\rm{old}} - \frac{q_s}{m_s} \left\langle\delta A^{\rm{h}}_{\|,\rm{old}} \right\rangle  \;\;, \\
\label{eq:pullback_df}
    & \delta f_{\rm{new}} = \delta f_{\rm{old}} +  \frac{q_s\left\langle\delta A^{\rm{h}}_{\|,\rm{old}} \right\rangle}{m_s}\frac{\partial f_{0s}}{\partial v_\|}
    \xrightarrow[f_{0s}=f_{M}]{\text{Maxwellian}}
     \delta f_{\rm{old}} -  \frac{ 2v_\|}{v_{\rm{t}s}^2}  \frac{q_s\left\langle\delta A^{\rm{h}}_{\|,\rm{old}} \right\rangle}{m_s} f_{0s} \;\;,
\end{align}
where variables with subscripts ``new'' and ``old'' refer to those after and before the pullback transformation, Eq.~(\ref{eq:pullback_df}) is the linearized equation for $\delta f$ pullback, which is from the general equation of the transformation for the distribution function $
    f_{\rm{old}} (u_{\| \rm{old}}) = f_{\rm{new}} \left(u_{\| \rm{new}} =u_{\| \rm{old}}- {q_s} \left\langle\delta A^{\rm{h}}_{\|,\rm{old}} \right\rangle/{m_s} \right)$. 
For the full $f$ scheme, only Eqs.~(\ref{eq:pullback_A}) and (\ref{eq:pullback_v}) are needed. 

Recently, the mixed variable-pullback scheme has been generalized by incorporating the non-ideal Ohm's law that recovers the traditional pure $v_\parallel$ scheme and the pure $p_\parallel$ scheme \cite{lu2025generalized}. 
Five schemes have been listed in the generalized pullback scheme, including the original one for which the symplectic part $\delta A_\|^{\rm{s}}$ is chosen to satisfy ideal Ohm's law involving the electrostatic scalar potential $\delta\Phi$ \cite{mishchenko2014pullback},
\begin{equation}
    \partial_t\delta A_\|^{\rm{s}}=-\partial_\|\delta\Phi\;\;,
\label{eq:ohm_law0}
\end{equation}
where the parallel derivative is defined as $\partial_\|=\boldsymbol{b}\cdot\nabla$, $\boldsymbol{b}=\boldsymbol{B}/B$, $\boldsymbol{B}$ is the equilibrium magnetic field. 
Another scheme is also implemented in TRIMEG-C1 with
\begin{equation}
    \partial_t\delta A_\|^{\rm{s}}=0\;\;.
\label{eq:ohm_law_dAdt0}
\end{equation}
Note that Eq.~\eqref{eq:ohm_law_dAdt0} is satisfied when solving the field and particle equations during each step before the pullback procedure, while at the end of each step, due to the pullback procedure, the value of $\delta A_\parallel^{\rm h}$ is shifted to $\delta A^{\rm s}_\|$ and thus, $\delta A^{\rm s}_\|$ becomes nonzero. 
%This scheme is particularly suitable for whole-volume simulations, since the ideal Ohm’s law can otherwise lead to a nonphysical component of $\delta A^{\rm s}_\|$ in regions with open magnetic field lines, where the field lines intersect the wall and thus $\partial_\parallel\delta\phi$ in Eq.~\eqref{eq:ohm_law0}. 
The scheme with the ideal Ohm's law is suitable for the MHD mode since the physics solution is very close to Eq.~\eqref{eq:ohm_law0}. For electrostatic modes, such as the ITG mode, Eq.\eqref{eq:ohm_law0} is less efficient. In such a case, Eq.~\eqref{eq:ohm_law_dAdt0} is more suitable, and no field equation for $\delta A_\parallel^{\rm s}$ needs to be solved. 
The two schemes are summarized in Tab.~\ref{tab:scheme12} while the more general treatment can be found in Tab. I in the previous work \cite{lu2025generalized}.
\begin{table}[h!]
\centering
\begin{tabular}{c c c c c}
\hline\hline 
& Equation & Pros\\
\hline 
Scheme I  & \eqref{eq:ohm_law0}  & Good approximation of MHD modes \\ \hline 
Scheme II & \eqref{eq:ohm_law_dAdt0} & No need to solve $\delta A^{\rm s}_\|$ equation; good for electrostatic modes \\ \hline 
\end{tabular}
\caption{\label{tab:scheme12} Two schemes implemented in TRIMEG-C1. The other three schemes can be found in \cite{lu2025generalized}. The pure $p_\parallel$ scheme solves the system with $\delta A^{\rm h}_\parallel=0$, which requires a large number of markers and a small time step size and thus is not considered in this work. The pure $v_\parallel$ scheme and the mixed variable-pullback scheme with the non-ideal Ohm's law require dedicated effort in implementing the non-ideal Ohm's law and will be studied separately. }
\end{table}

\section{Piecewise Field-aligned finite element method in triangular meshes}
\label{sec:numeric_scheme}
\subsection{Traditional 3D finite element method in TRIMEG-C1}
\label{subsec:FEM3d_traditional}
The traditional 3D finite element method using $C^1$ triangular finite elements  for the poloidal plane and cubic B-splines in toroidal direction is developed in TRIMEG-C1 as reported in our previous work \cite{lu2024gyrokinetic} and is summarized as follows. 
In the 3D simulation domain $\Omega$, there are $N_{\rm vert}$ vertices in each poloidal plane and $N_\phi$ grids in the toroidal direction. Each poloidal plane is decomposed into $N_{\rm tria}$ triangles. Each triangle is mapped to a reference triangle whose three vertices are $(\xi,\eta)=(0,0),(1,0),(0,1)$, where $\xi,\eta$ are the reference coordinates. The global coordinates and the reference coordinates can be mapped to each other for a given triangle $T_I$, namely
\begin{eqnarray}
% \label{eq:RZ2xieta}
%     \xi=\xi_{I}(R,Z)\;\;,\;\;\eta=\eta_{I}(R,Z)\;\;, \\
\label{eq:xieta2RZ}
    R=R_I(\xi,\eta)\;\;,\;\;Z=Z_I(\xi,\eta)\;\;,
\end{eqnarray}
where $(\xi,\eta)$ is in the reference triangle, \textit{i.e.} $\xi\in[0,1]$, $\eta\in[0,1]$, $\xi+\eta\in[0,1]$. For linear mappings as adopted in this work, we have $R=(R_{2,I}-R_{1,I})\xi+(R_{3,I}-R_{1,I})\eta+R_{1,I}$, $Z=(Z_{2,I}-Z_{1,I})\xi+(Z_{3,I}-Z_{1,I})\eta +Z_{1,I}$,  where $(R_{I,I},Z_{1,I})$, $(R_{2,I},Z_{2,I})$ and $(R_{3,I},Z_{3,I})$ are the coordinates of the three vertices of triangle $i$. 
% In the following, for the sake of simplicity, the subscript `i' is omitted in $\xi$ and $\eta$. 
Along the toroidal direction, regular grids and the cubic B-spline are adopted. In each 2D triangle, $N_{\rm bas}=18$ basis functions $\Lambda_i(\xi,\eta)$, where $i=1,\ldots,N_{\rm bas}$, are defined in the reference coordinate $(\xi,\eta)$ \cite{lu2024gyrokinetic}. 

By applying the continuity condition to the value, the first and second derivatives of the represented function, the degree of freedom in each poloidal plane is reduced from $18N_{\rm tria}$ to $6N_{\rm vert}$ before applying the boundary condition. 
Correspondingly, the three-dimensional variable $Y(R,\phi,Z)$ is written as follows,
\begin{eqnarray}
\label{eq:Y3d_traditional}
	Y(R,\phi,Z) &=& \sum_{i=1}^{N_{\rm vert}}\sum_{j=1}^6\sum_{k=1}^{N_\phi} \widehat\Lambda_{ij}(R_I(\xi,\eta),Z_I(\xi,\eta)) \Gamma_k(\phi) \widehat{Y}_{i,j,k}  \;\;,
\end{eqnarray}
%(Eric: add eta,xi-R,Z; in a given triangle): done 
where $\widehat{\Lambda}_{i,j}(R_I(\xi,\eta),Z_I(\xi,\eta))$ are linear combinations of $\Lambda_J(\xi,\eta)$ for $(R,Z)\in T_I$ (Eq. (73) of \cite{lu2024gyrokinetic}). Here the index $J$ between 1 and $N_{bas}$ depends on the vertex $i$ and the triangle $I$ through the mapping.
Due to this construction, the support of  $\widehat{\Lambda}_{i,j}$ corresponds to all the triangles with vertex $i$ and $\widehat{\Lambda}_{i,j}$ is in $C^1$.
%is the linear combination of $\Lambda$ satisfying the continuity, 
On the other hand $\Gamma_k(\phi)$ is the cubic spline basis function, and $N_\phi$ is the grid point number in the toroidal direction. Since there is a one-to-one mapping between the 
reference coordinates $(\xi,\eta)$ and the global coordinates $(R,Z)$ in each triangle as indicated by Eq.~\eqref{eq:xieta2RZ}, Eq.~\eqref{eq:Y3d_traditional} can be also written as
\begin{eqnarray}
\label{eq:Y3dRZ_traditional}
	Y(R,\phi,Z) &=&  \sum_{i=1}^{N_{\rm vert}}\sum_{j=1}^6\sum_{k=1}^{N_\phi} \widehat\Lambda_{ij}(R,Z) \Gamma_k(\phi) \widehat{Y}_{i,j,k}  \;\;.
\end{eqnarray}
The weak form of the equation to be solved, needed for a finite element formulation, can be obtained by multiplying it by the test function $ \widehat\Lambda_{i'j'}(R,Z) \Gamma_{k'}(\phi)$ and calculating the volume integral in the simulation domain. The coefficients $\widehat{Y}_{i,j,k}$ can be obtained from the weak form and are used for calculating the field at the particle location for pushing particles. More details can be found in the previous work \cite{lu2024gyrokinetic}. 

\subsection{Piecewise field-aligned finite element method in $(R,\phi,Z)$ coordinates}
\label{subsec:PFAFEM}
The piecewise field-aligned finite element method (PFAFEM) has been proposed to study the multi-harmonic nonlinear physics in tokamak plasmas with increased computational efficiency. 
Previously, it has been implemented in the electrostatic particle model for the core plasma in the tokamak coordinate $(r,\phi,\theta)$, where $r$, $\phi$, and $\theta$ are the radial-like, toroidal-like, and poloidal-like coordinates \cite{lu2025piecewise}. In this work, the PFAFEM has been extended and implemented in cylindrical coordinates for electromagnetic whole-volume simulations. The two key features of the PFAFEM still hold: 
\begin{enumerate}
    \item The computational grids are aligned in a traditional pattern without any shift.
    \item The finite element basis functions are defined on piecewise field-aligned coordinates, with each basis function being continuous along the magnetic field line. 
\end{enumerate}
The previous treatment of PFAFEM in $(r,\phi,\theta)$ is summarized as follows \cite{lu2025piecewise}.
We first defined the $k$'th subdomain as 
\begin{eqnarray}
\label{eq:subdomain_define}
    \phi-\phi_k\in[-N_{\rm bsp}\Delta\phi/2,N_{\rm bsp}\Delta\phi/2]
\end{eqnarray} where $\phi_k=(k-1)\Delta\phi$ is the coordinate of the toroidal grid point, $\Delta\phi$ is the toroidal grid spacing, and $N_{\rm bsp}$ is the order of the B-Spline (for cubic B-spline, $N_{\rm bsp}=4$).
In $(r,\phi,\theta)$ coordinates, the safety factor is $q(r,\theta)\equiv{\bf B}\cdot\nabla\phi/({\bf B}\cdot\nabla\theta)$. The piecewise field-aligned local coordinate $\eta_{k}$ is defined in each toroidal subdomain centered at $\phi_k$ grid as 
$
\eta_{k}(r,\theta,\phi)= \theta-\int_{\phi_k}^\phi {\rm d}\phi'{1}/{q(r,\theta'(r,\phi'),\phi')}
$, 
where the integral is along the magnetic field line and the safety factor $q=q(r,\theta',\phi')$, $\theta'$ is determined by following the magnetic field while varying $\phi'$, namely, $d\theta'/d\phi'=1/q(r,\theta',\phi')$, $\phi_k$ and $\phi$ denote the starting and end points of the integral, respectively. For the straight field line coordinates $r,\bar\theta,\bar\phi$, we have
$\eta_{k}(r,\bar\theta,\bar\phi)=\bar\theta-({\bar\phi-\bar\phi_k})/{\bar q}$, 
where the safety factor $\bar q=\bar q(r)$. More details can be found in the previous work \cite{lu2025piecewise}. 

In this work, for the whole volume simulation, we construct two field-aligned coordinates $R_{\rm FA}$ and $Z_{\rm FA}$, by tracing the equilibrium magnetic field lines. 
The subdomain is defined in Eq.~\eqref{eq:subdomain_define}. 
The piecewise field-aligned local coordinate $R_{{\rm FA}k}$ and $Z_{{\rm FA}k}$ are defined in each toroidal subdomain centered at $\phi_k$ grid as follows,
\begin{eqnarray}
\label{eq:eta_integral}
    R_{{\rm FA},k}(R,Z,\phi)= R-\int_{\phi_k}^\phi {\rm d}\phi'\frac{{\bf b}\cdot\nabla R'}{{\bf b}\cdot\nabla\phi'}  \;\;, \\
    Z_{{\rm FA},k}(R,Z,\phi)= Z-\int_{\phi_k}^\phi {\rm d}\phi'\frac{{\bf b}\cdot\nabla Z'}{{\bf b}\cdot\nabla\phi'}  \;\;, 
\end{eqnarray}
where the integral is along the magnetic field line. 
Numerically,  $R_{{\rm FA}k}$ and $Z_{{\rm FA}k}$  are calculated on the 3D grids $(R_I,Z_J,\phi_K)$ and stored as 3D matrices $R_{{\rm FA},k,IJK}$ and $Z_{{\rm FA},k,IJK}$. The values of $R_{{\rm FA},k}$ and $Z_{{\rm FA},k}$  on each grid point $(R_I,Z_J,\phi_K)$ is calculated by solving the characteristic line from $\phi_K$ to the middle toroidal position $\phi_k$ of the subdomain, namely, the $\phi$ coordinate of toroidal grids, using the Runge-Kutta 4th order scheme, 
\begin{eqnarray}
    \frac{dR_{{\rm FA},k}}{d\phi} = \frac{{\bf b }\cdot\nabla R}{{\bf b }\cdot\nabla \phi}\;\;, \\
    \frac{dZ_{{\rm FA},k}}{d\phi} = \frac{{\bf b }\cdot\nabla Z}{{\bf b }\cdot\nabla \phi}\;\;, 
\end{eqnarray}
which is called ``backward field line tracing'' since the direction of the integral is from $\phi$ back to the reference toroidal location $\phi_k$. 
Then with  $R_{{\rm FA},k,IJK}$ and $Z_{{\rm FA},k,IJK}$, the values of $R_{{\rm FA}k}$ and $Z_{{\rm FA}k}$ at a given position $(R,Z,\phi)$ can be obtained by B-spline interpolation.  

The piecewise field-aligned finite element basis function is defined in $(R_{{\rm FA},k},\phi,Z_{{\rm FA},k})$. A perturbed field variable $Y$ is represented as the summation of these field-aligned finite element basis functions as follows,
\begin{eqnarray}
\label{eq:Y3dRZFA_PFAFEM}
    Y(R,\phi,Z) =\sum_{i=1}^{N_{\rm vert}}\sum_{j=1}^6\sum_{k=1}^{N_\phi} \widehat\Lambda_{i,j}(R_{{\rm FA},k},Z_{{\rm FA},k})\Gamma_k(\phi)\widehat{Y}_{i,j,k} \;\;.
\end{eqnarray}
Compared with the traditional 3D finite element method in Eq.~\eqref{eq:Y3dRZ_traditional}, the  difference is that using PFAFEM, $\widehat\Lambda$ is defined in $(R_{{\rm FA},k},Z_{{\rm FA},k})$ instead of $(R,Z)$. 
The interpolation of the perturbed field at the particle location (``gathering'')  can be performed in $(R_{{\rm FA},k},\phi,Z_{{\rm FA},k})$ using Eq.~\eqref{eq:Y3dRZFA_PFAFEM}. The derivative is calculated using the chain rule,
\begin{eqnarray}
\label{eq:dY3dRFA_PFAFEM}
    \frac{\partial}{\partial R}Y(R,\phi,Z) &=&\sum_{i=1}^{N_{\rm vert}}\sum_{j=1}^6\sum_{k=1}^{N_\phi} \left( 
     \frac{\partial R_{\rm FA}}{\partial R}\frac{\partial}{\partial R_{{\rm FA},k}}\widehat\Lambda_{i,j} 
    + \frac{\partial Z_{\rm FA}}{\partial R}\frac{\partial}{\partial Z_{{\rm FA},k}}\widehat\Lambda_{i,j}
    \right)\Gamma_k(\phi)\widehat{Y}_{i,j,k} \;, \\
\label{eq:dY3dZFA_PFAFEM}
    \frac{\partial}{\partial Z}Y(R,\phi,Z) &=&\sum_{i=1}^{N_{\rm vert}}\sum_{j=1}^6\sum_{k=1}^{N_\phi} \left( 
     \frac{\partial R_{\rm FA}}{\partial Z}\frac{\partial}{\partial R_{{\rm FA},k}}\widehat\Lambda_{i,j} 
    + \frac{\partial Z_{\rm FA}}{\partial Z}\frac{\partial}{\partial Z_{{\rm FA},k}}\widehat\Lambda_{i,j}
    \right)\Gamma_k(\phi)\widehat{Y}_{i,j,k} \;,  \\
\label{eq:dY3dphiFA_PFAFEM}
    \frac{\partial}{\partial \phi}Y(R,\phi,Z) &=&\sum_{i=1}^{N_{\rm vert}}\sum_{j=1}^6\sum_{k=1}^{N_\phi} \left[\left( 
     \frac{\partial R_{\rm FA}}{\partial \phi}\frac{\partial}{\partial R_{{\rm FA},k}}\widehat\Lambda_{i,j} + \frac{\partial Z_{\rm FA}}{\partial \phi}\frac{\partial}{\partial Z_{{\rm FA},k}}\widehat\Lambda_{i,j}
    \right)\Gamma_k(\phi)\right.\nonumber\\
    &+&
    \left.
    \widehat{\Lambda}_{i,j}\frac{\partial}{\partial\phi}\Gamma_k(\phi)\right]\widehat{Y}_{i,j,k} \;, 
\end{eqnarray}
where we have omitted the arguments in $\widehat{\Lambda}=\widehat{\Lambda}(R_{{\rm FA},k},Z_{{\rm FA},k})$. In the implementation, Eqs.~\eqref{eq:dY3dRFA_PFAFEM}--\eqref{eq:dY3dphiFA_PFAFEM} are adopted to  calculate the spatial derivatives of $\delta\Phi$ and $\delta A_\parallel$ in the $R,\phi,Z$ directions to push particles in $(R,\phi,Z)$ coordinates.

The calculation of the perturbed density and parallel current (``scattering'') is obtained by the projection operator from particles to the field-aligned finite element basis functions, as to be introduced in Section \ref{subsec:solvers}.

%In addition, the matrices for the field equations are calculated in $(R_{{\rm FA},k},\phi,Z_{{\rm FA},k})$. PFAFEM is consistent with the theoretical studies of the wave packet in tokamak plasmas \cite{lu2012theoretical}.
The partition of unity is obtained readily for the traditional 3D scheme. In the poloidal plane,  noting that $\Lambda_1$, $\Lambda_7$, and $\Lambda_{13}$ are associated with the values of the function while the other basis functions are associated with the derivatives, the partition of unity is $\Lambda_1+\Lambda_7+\Lambda_{13}=1$. 
To prove it, first, write each term in a symmetric way, 
$\Lambda_1
= 10\lambda^3 - 15\lambda^4 + 6\lambda^5
+ 30\lambda^2 \xi \eta (\xi + \eta)$,
$\Lambda_7
= 10\xi^3 - 15\xi^4 + 6\xi^5
+ 15\xi^2 \eta^2 \lambda$, 
$\Lambda_{13}
= 10\eta^3 - 15\eta^4 + 6\eta^5
+ 15\xi^2 \eta^2 \lambda $. 
Then the summation yields
$\Lambda_1 + \Lambda_7 + \Lambda_{13}
= 10(\lambda^3 + \xi^3 + \eta^3)
- 15(\lambda^4 + \xi^4 + \eta^4) + 6(\lambda^5 + \xi^5 + \eta^5)
+ 30\lambda^2 \xi \eta (\xi + \eta)
+ 30\xi^2 \eta^2 \lambda$ \cite{lu2024gyrokinetic}. 
After expanding and collecting terms, using $\lambda = 1 - \xi - \eta $, the expression reduces to the constant $1$. 
Then also noting Eqs.~(73) and (74) in our previous work \cite{lu2024gyrokinetic}, we have
\begin{eqnarray}
    \widehat\Lambda_{i_1,1}+\widehat\Lambda_{i_2,1}+\widehat\Lambda_{i_3,1}=1\;\;,
\end{eqnarray}
where $i_1,i_2,i_3$ are the three vertices of a triangle, and $\widehat\Lambda_{i_1,1}=\Lambda_1$, $\widehat\Lambda_{i_1,2}=\Lambda_7$, $\widehat\Lambda_{i_1,3}=\Lambda_{13}$. 

%Eric: Add lambda1--hatlambdaI1 (also add subdomain: move to the front): Done 

For traditional 3D FEM, the partition of unity is proved as follows,
\begin{eqnarray}
\label{eq:partition1traditional}
    \sum_{k=1}^{N_\phi}\left[\widehat\Lambda_{i_1,1}(R_p,Z_p)+\widehat\Lambda_{i_2,1}(R_p,Z_p)+\widehat\Lambda_{i_3,1}(R_p,Z_p)\right] \Gamma_k(\phi_p) =\sum_{k=1}^{N_\phi}\Gamma_k(\phi_p)  =1\;\;,
\end{eqnarray}
where $(R_p,\phi_p,Z_p)$ denote the particle location.

For the piecewise field-aligned FEM, 
\begin{eqnarray}
\label{eq:partition1FA}
    \sum_{k=1}^{N_\phi} \Gamma_k(\phi_p)\left[\widehat\Lambda_{i_1,1}(R_{p,k},Z_{p,k})+\widehat\Lambda_{i_2,1}(R_{p,k},Z_{p,k})+\widehat\Lambda_{i_3,1}(R_{p,k},Z_{p,k})\right] =\sum_{k=1}^{N_\phi}\Gamma_k(\phi_p)  =1\;\;,
\end{eqnarray}
since $\widehat\Lambda_{i_1,1}(R_{p,k},Z_{p,k})+\widehat\Lambda_{i_2,1}(R_{p,k},Z_{p,k})+\widehat\Lambda_{i_3,1}(R_{p,k},Z_{p,k})=1$ applies for a given value of $k$ and the particle location. 
The partition of unity has also been discussed and proved previously in the mesh-free scheme \cite{mcmillan2017partially}.
It can be readily demonstrated that Eqs.~\eqref{eq:partition1traditional} and~\eqref{eq:partition1FA} ensure charge (particle-number) conservation in the projection operator in Eqs.~\eqref{eq:project_NFA} and \eqref{eq:project_FA}.

The constructed $(R_{{\rm FA},k}, Z_{{\rm FA},k})$ grids are demonstrated in Fig.~\ref{fig:RZ_pfafem}. It corresponds to the TCV-X21 \cite{oliveira2022validation,body2022development,ulbl2023influence} case used for the numerical studies in Section \ref{sec:results}. The coordinates $(R_{{\rm FA},k}, Z_{{\rm FA},k})$ are evaluated on the sub-grids in each toroidal subdomain. Then $(R_{{\rm FA},k}$ and $Z_{{\rm FA},k})$ are numerically represented as functions of $(R,Z,\phi)$. The grid in $(R_{{\rm FA},k}, Z_{{\rm FA},k})$ is distorted more strongly as the toroidal location is farther away from the mid location of the toroidal subdomain, as demonstrated in Fig.~\ref{fig:RZ_pfafem}. This also demonstrates a key advantage of PFAFEM in mitigating grid distortion, which becomes severe when global field-aligned coordinates are used. By constructing \((R_{{\rm FA},k}, Z_{{\rm FA},k})\) locally within each toroidal subdomain \(\phi - \phi_k \in [-N_{\rm bsp}\Delta\phi/2, N_{\rm bsp}\Delta\phi/2]\), the resulting distortion is significantly reduced compared to that over the full toroidal domain \([0, 2\pi]\), where magnetic shear has a more pronounced effect on the grid distortion. Furthermore, via a higher number of toroidal finite elements, the grid distortion can be reduced to the required level. 

%%%%%%%%%%%%%%%%%%  %%%%%%%%%%%%%%%%%%%
%\begin{figure}[t]
%\centering
%\includegraphics[width=0.8\textwidth]{./figures/eqdsk_tcvx21.eps}
%\caption{The poloidal flux function, the safety factor, and the pressure profiles of the TCV-X21 case.}\label{fig:RZ_pfafem}
%\end{figure}

%%%%%%%%%%%%%%%%%%  %%%%%%%%%%%%%%%%%%%
\begin{figure*}[t]
\centering
\includegraphics[width=0.8\textwidth]{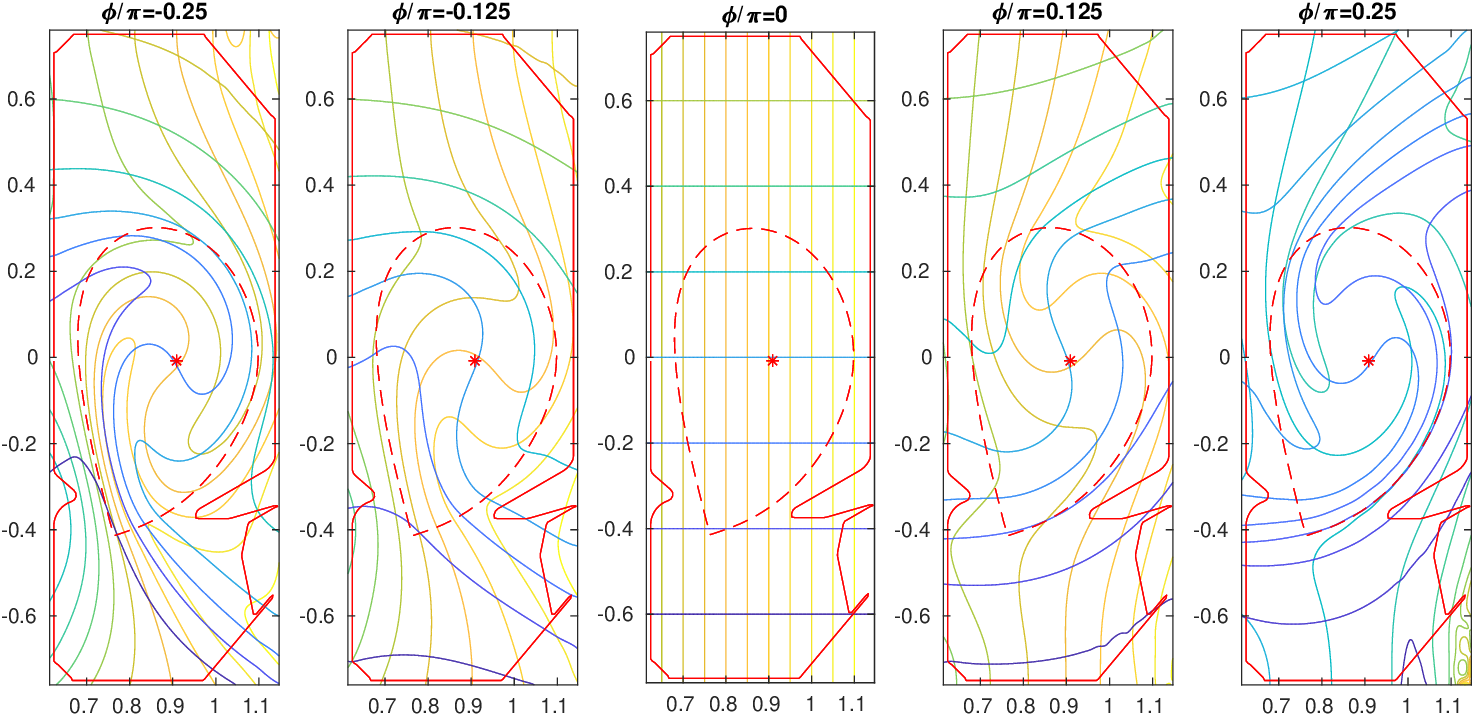}
\caption{The grid of the piecewise field-aligned coordinates $(R_{\rm FA},Z_{\rm FA})$ (Section \ref{subsec:PFAFEM}).}
\label{fig:RZ_pfafem}
\end{figure*}

\subsection{The 3D traditional and the PFAFEM field solvers}
\label{subsec:solvers}
Four field equations are solved: the quasi-neutrality equation, Amp\`ere's law, the iterative Amp\`ere equation, and Ohm's law. We first discuss the general form of the field equation and later demonstrate the four field equations separately.
%\subsubsection{General form of the field equation}
The general form of the field equation is
\begin{eqnarray}
\label{eq:field_equation_general}
    M_L\boldsymbol{\cdot} {\bf y} =   {\bf b}+M_R\boldsymbol{\cdot} {\bf c}\;\;,
\end{eqnarray}
where $M_L$ and $M_R$ are partial differential operators, $\bf b$ and $\bf c$ are known vectors, $\bf y$ is the vector to be solved, and the physics variable $b$ corresponding to the vector $\bf b$ is calculated from the particle distribution function,
\begin{eqnarray}\label{eq:b_integral}
    b=\int {\rm d}^3 v \delta f Q(\boldsymbol{v}) \;\;,
\end{eqnarray}
where $Q(\boldsymbol{v})$ is a function of the velocity (for the quasi-neutrality equation, $Q(\boldsymbol{v})=1$; for Amp\`ere's law, $Q(\boldsymbol{v})=v_\|$).
The corresponding general matrix form of the field equations is 
\begin{eqnarray}
\label{eq:mat_poisson}
    \bar{\bar{M}}_{L,ii',jj',kk'} \boldsymbol{\cdot} Y_{i'j'k'}
    &=&   B^{i,j,k} + \bar{\bar{M}}_{R,ii',jj',kk'} \boldsymbol{\cdot} C_{i'j'k'}\;\;,\\
    \label{eq:field3dfem}
    \bar{\bar{M}}_{L/R,ii',jj',kk'}
    &=& \int {\rm d} R\, {\rm d}Z\, {\rm d}\phi\, J \Upsilon_{ijk} 
    M_{L/R}\Upsilon_{i'j'k'}\;\;, 
\end{eqnarray}
where $Y_{i'j'k'}$ is the field variable to be solved by the linear solver, $B^{i,j,k}$ is from the markers using the projection operator, $C_{i'j'k'}$ is the known field variable, $\bar{\bar{M}}_{L,ii',jj',kk'}$ and $\bar{\bar{M}}_{R,ii',jj',kk'}$ are the matrices on the left- and right-hand sides, respectively. The 3D basis functions $\Upsilon_{ijk}$ and $\Upsilon_{i'j'k'}$ denote those in  $(R,\phi,Z)$ coordinate for the traditional solver or those in $(R_{\rm FA},\phi,Z_{\rm FA})$ coordinate for the PFAFEM solver. We will list them separately below.

For the traditional 3D FEM solver, 
\begin{eqnarray}
    \label{eq:Nijk3dfem}
    &&\Upsilon_{ijk}=\Upsilon_{ijk}(R,Z,\phi) = \widehat\Lambda_{ij}(R,Z)\Gamma_k(\phi)\;\;, \\
    \label{eq:project_NFA}
    &&B^{i,j,k}
    =
    \sum_s C_{{\rm p2g},s}\sum_{p=1}^{N_{\rm mark}} w_p Q(\boldsymbol{v}_p) \Upsilon_{ijk}(R_{p},Z_{p},\phi_p) \;\;, 
\end{eqnarray}
where the conversion factor in the projection operator $C_{{\rm p2g},s}=-\Bar{q}_s \langle n\rangle_V V_{\rm tot}/N_{\rm mark}$. 

The matrix is calculated using a modified Gauss quadrature. The general form of the matrix  in $\bar{\bar{M}}_{L/R,ii',jj',kk'}$ contains the volume integral after the integral by parts when $M_{L/R}$ is a second (or higher) order differential operator,
\begin{eqnarray}
    m_{ii',jj',kk'}=\int {\rm d} r\, {\rm d}\theta\, {\rm d}\phi\, J L_1 \Upsilon_{ijk} 
    L_2\Upsilon_{i'j'k'}\;\;,
\end{eqnarray}
where $L_1$ and $L_2$ are differential operators associated to $M_{L/R}$. For example, for $M_{L/R}=\partial^2/\partial R^2$, we have $L_1=-\partial/\partial R$ and $L_2=\partial/\partial R$. Using the modified Gauss quadrature in prisms
%(Eric: check the whole manuscript ): done
\begin{eqnarray}
    m_{ii',jj',kk'} = \sum_{I=1}^{I_{\rm max}}\sum_{J=1}^{J_{\rm max}} J L_1 \Upsilon_{ijk} (R_{I},\phi_J,Z_I)
    L_2\Upsilon_{i'j'k'}(R_{I},\phi_J,Z_I)\;\;,
\end{eqnarray}
where $I$ and $J$ denote the indices of the quadrature points in the reference coordinate $(\xi,\eta)$ and in the toroidal direction, respectively.

The toroidal filter can be applied to single-$n$ simulations in the traditional 3D solver, following the previous approach for deriving the 2D filter \cite{lu2023full}, and then reduced to the 1D toroidal filter. The toroidal harmonic is kept as follows, 
\begin{eqnarray}
    b(R,\phi,Z)=b_n(R,Z)\mathrm{e}^{\mathrm{i}n\phi}+\rm{c.c.} \;\;,
\end{eqnarray}
where $\rm c.c.$ denotes the complex conjugate $b_{-n}(R,Z)\mathrm{e}^{-\mathrm{i}n\phi}$ for nonzero $n$ while for $n=0$, $\rm c.c.$ is zero. 
%Substituting Eq.~\eqref{eq:b_integral} to $b_n=\int_0^{\phi_{\rm wid}}{\rm d}\phi \mathrm{e}^{-\mathrm{i}n\phi}b(R,\phi,Z)/\phi_{\rm wid}$, we obtain
%\begin{eqnarray}
%    b_n(R,Z)&=&\frac{C_{{\rm p2g},s}}{\phi_{\rm wid}}\sum_{p=1}^{N_{\rm mark}} w_p Q({\boldsymbol{v}})\mathrm{e}^{-\mathrm{i}n\phi_p}\;\;,
%\end{eqnarray}
%where $\phi_{\rm wid}$ is the width of the simulation domain in the toroidal direction and is typically chosen as $\phi_{\rm wid}=2\pi/n$ for single-$n$ simulations. 
To apply the toroidal Fourier filter to the coefficient $B^{i,j,k}$, we first calculate the Fourier component as follows, 
%project $b$ to $(\widehat{\Lambda},\mathrm{e}^{inp})$,
\begin{eqnarray}
    b_{ijn}&=&\frac{C_{{\rm p2g},s}}{\phi_{\rm wid}}\sum_{p=1}^{N_{\rm mark}} w_p Q({\boldsymbol{v}})\mathrm{e}^{-\mathrm{i}n\phi_p}\widehat{\Lambda}_{ij}(R_p,Z_p)\;\;.
\end{eqnarray}
Then Eq.~\eqref{eq:Nijk3dfem} is replaced with
\begin{eqnarray}
    B^{i,j,k}_{\rm filter}
    = \sigma_n {\rm Re}(T_{n,k} \mathrm{e}^{\mathrm{i}n\phi_k})\Delta\phi 
   b_{ijn}
    \;\;,
    %\frac{C_{{\rm p2g},s}}{\phi_{\rm wid}}\sum_{p=1}^{N_{\rm mark}} w_p Q({\boldsymbol{v}})\mathrm{e}^{-\mathrm{i}n\phi_p}\widehat{\Lambda}_{ij}(R_p,Z_p)
\end{eqnarray}
where $\sigma_n=2$ for $n\neq0$ and $\sigma_n=1$ for $n=0$, $T_{n,k}=[6+2\cos(2n_{\rm eff})-8\cos(n_{\rm eff})]/n_{\rm eff}^4=[2\sin(n_{\rm eff}/2)/n_{\rm eff}]^4$ for $n\neq0$ and $T_{n,k}=1$ for $n=0$, $n_{\rm eff}=n\Delta\phi$. 
    
For the 3D field-aligned FEM solver, 
\begin{eqnarray}
    &&\Upsilon_{ijk}=\Upsilon_{ijk}(R_{{\rm FA} ,k},Z_{{\rm FA} ,k},\phi ) =\widehat\Lambda_{ij}(R_{{\rm FA},k},Z_{{\rm FA},p,k})\Gamma_k(\phi ) \;\;,\\
    \label{eq:Nijk3dpfafem}
    &&B^{i,j,k}
    =
    \sum_s C_{{\rm p2g},s}\sum_{p=1}^{N_{\rm mark}} w_p Q(\boldsymbol{v}_p) \Upsilon_{ijk}(R_{{\rm FA},p,k},Z_{{\rm FA},p,k},\phi_p) \;\;.
    \label{eq:project_FA}
\end{eqnarray}

The matrix is calculated using the modified Gauss quadrature and the mixed particle-wise-basis-wise scheme \cite{lu2025piecewise} but expressed in $(R_{\rm FA},\phi,Z_{\rm FA})$ coordinates. 
The integral points $(R_{{\rm FA},I,k},\phi_J,Z_{{\rm FA},I,k})$ are determined in the basis function $\Upsilon_{ijk}$ (grid-wise), and the corresponding $(R_{I,k},\phi_J,Z_{I,k})$ are calculated to determine the integral points in $\Upsilon_{i'j'k'}$ (particle-wise). The matrix element is calculated as follows, 
\begin{eqnarray}
    m_{ii',jj',kk'} = \sum_{I=1}^{I_{\rm max}}\sum_{J=1}^{J_{\rm max}} J L_1 \Upsilon_{ijk} (R_{{\rm FA},I,k},\phi_J,Z_{{\rm FA},I,k})
    L_2\Upsilon_{i'j'k'}(R'_{{\rm FA},I,k'},\phi_J,Z'_{{\rm FA},I,k'})\;\;,
\end{eqnarray}
where $R'_{{\rm FA},I,k'}$ and $Z'_{{\rm FA},I,k'}$ are determined by
\begin{eqnarray}
    R(R'_{{\rm FA},I,k'},\phi_J-\phi_{k'},Z'_{{\rm FA},I,k'})=R(R_{{\rm FA},I,k},\phi_J-\phi_{k},Z_{{\rm FA},I,k})\;\;, \\
    Z(R'_{{\rm FA},I,k'},\phi_J-\phi_{k'},Z'_{{\rm FA},I,k'})=Z(R_{{\rm FA},I,k},\phi_J-\phi_{k},Z_{{\rm FA},I,k})\;\;, 
\end{eqnarray}
since each integral point is identical regardless of the coordinate system. 

\subsection{Weak form of the field equations}
\label{subsec:weak_form}
To express the weak form of the field equations, we first give the normalized forms. The normalized quasi-neutrality equation is, 
\begin{eqnarray}
\label{eq:poisson_normalized}
    \bar\nabla_\perp \cdot \left(G_{\rm P}\bar\nabla_\perp\delta \bar{\Phi}\right)=-C_{\rm P}\delta\bar N\;\;, \;\;
    G_{\rm P}=\sum_s \frac{n_{0s}}{n_{\rm ref}} \bar{m}_s \left(\frac{B_{\rm N}}{B}\right)^2 \;\;,
\end{eqnarray} 
where $C_{\rm P}=1/\bar\rho_{\rm{ref}}^2$, $\delta\bar{N}_s=\delta n/n_{\rm ref}$, $\delta\bar{N}=\sum_s\bar q_s\delta\bar{N}_s$.

The normalized Amp\`ere's law and the iterative scheme corresponding to Eqs.~(\ref{eq:ampere_h0})--(\ref{eq:A2ndavg_fullf}),
\begin{eqnarray}
\label{eq:ampere_normalize0}
    &&\left(\bar\nabla^2_\perp-\sum_s\frac{\bar{q}_s^2}{\bar m_s}C_{\mathrm A}\right)\delta \bar A_{\|,0}^{\rm{h}} 
    = -\bar\nabla^2_\perp \delta\bar A_{\|,0}^{\rm{s}} - C_{\mathrm A} \delta \bar J_{\|} \;\;, 
    \\
\label{eq:ampere_normalize1}
    &&\left(\bar\nabla^2_\perp-\sum_s\frac{\bar{q}_s^2}{\bar m_s}C_{\mathrm A}\right)\delta\bar A_{\|,I}^{\rm{h}} 
    = -\sum_s\frac{\bar{q}_s^2}{\bar m_s}C_{\mathrm A}\delta\bar A_{\|,I-1}^{\rm{h}} 
    + \bar{G} \delta\bar A_{\|,I-1}^{\rm{h}} 
    \\
\label{eq:skin_df_normalize}
    &&\bar{G}\delta\bar A_{\|,I-1}^{\rm{h}} =C_{\mathrm A}\frac{N_{0s} \bar{q}_s^2}{\bar{T}_s}\sum_{p=1}^N 2\bar v_{\parallel,p}^2 
    \int {\rm d}z^6 w_p\delta(\Tilde{\mathbf{R}}_p)\langle\delta\bar A_{\|,I-1}^{\rm{h}}\rangle   {\text{ for $\delta f$}}\;\;,
    \\
\label{eq:skin_fullf_normalize}
    &&\bar{G}\delta\bar A_{\|,I-1}^{\rm{h}} =C_{\mathrm A}\frac{N_{0s} \bar{q}_s^2}{\bar m_s}\sum_{p=1}^N 
    \int {\rm d}z^6 p_{p,{\rm tot}}\delta(\Tilde{\mathbf{R}}_p) 
    \langle\delta\bar A_{\|,I-1}^{\rm{h}}\rangle   {\text{ for full $f$}} \;\;,
\end{eqnarray}
where {$\Tilde{\boldsymbol{R}}_p={\boldsymbol{R}_p+\mathbf{\rho}_p-\mathbf{x}}$, $\mathbf{\rho}_p=m_s\mathbf{v}_{\perp,p}/(q_sB)$, $\mathbf{x}$ denotes the particle location,} $C_{\mathrm A} =\beta_{\rm ref}/\rho_{\rm ref}^2$, $n_{\rm ref}$ is the reference density and is chosen as the electron density on magnetic axis, and $\delta\Bar{J}_\|=\delta j_\|/(ev_{\rm N} n_{\rm ref})$.

The Ohm's law is
\begin{equation}
\label{eq:ohm_law_scheme12}
    \partial_t\delta \bar{A}_\|^{\rm{s}}=    
\begin{cases}
-\partial_\|\delta\bar\Phi \;\;,   & \text{scheme I}, \\
0 \;\;,  & \text{scheme II}. 
\end{cases}
\end{equation}

The weak form is expressed as follows
\begin{eqnarray}
\label{eq:mat_poisson}
    {\rm (\ref{eq:poisson_normalized})} \Rightarrow
    \bar{\bar{M}}_{\mathrm{P},L,ii',jj',kk'} \boldsymbol{\cdot}\delta\Phi_{i'j'k'}
    &&= C_{\rm P} \int {\rm d}R{\rm d}Z{\rm d}\phi\, J\delta N^{i,j,k}\Upsilon_{ijk} \;\;,  \\
\label{eq:mat_ampere0}
    {\rm (\ref{eq:ampere_normalize0})} \Rightarrow 
    \bar{\bar{M}}_{\mathrm{A},L,ii',jj',kk'} \boldsymbol{\cdot}\delta A^{\rm{h}}_{i'j'k',I=0} 
    &&= \bar{\bar{M}}_{\mathrm{A},R,ii',jj',kk'} \boldsymbol{\cdot}\delta A^{\rm{s}}_{i'j'k'}+ C_{\mathrm A} \delta J^{i,j,k}_\| \;\;,  \\
\label{eq:mat_ampere1}
    {\rm (\ref{eq:ampere_normalize1})}\Rightarrow
    \bar{\bar{M}}_{\mathrm{it},L,ii',jj',kk'} \boldsymbol{\cdot}\delta A^{\rm h}_{i'j'k',I+1} 
    &&= \bar{\bar{M}}_{\mathrm{it},R,ii',jj',kk'} \boldsymbol{\cdot}\delta A^{\rm h}_{i'j'k',I} 
    +\sum_s \frac{1}{d_{s}^2}\overline{\langle\delta A^{\rm h}_{i,j,k,I} \rangle} \;, \\
\label{eq:mat_ohm}
    {\rm (\ref{eq:ohm_law_scheme12})} \Rightarrow
    \bar{\bar{M}}_{\mathrm{Ohm},L,ii',jj',kk'} \boldsymbol{\cdot}\partial_t \delta A^{\rm{s}}_{i'j'k'} 
    &&= 
    \bar{\bar{M}}_{\mathrm{Ohm},R,ii',jj',kk'} \boldsymbol{\cdot} \delta\Phi_{i'j'k'}  \;\; \text{ for scheme I} \;\;.
\end{eqnarray}
{For scheme II, we do not need to solve any Ohm's law. 
Equations (\ref{eq:mat_poisson})--(\ref{eq:mat_ohm}) are solved numerically which provides the perturbed fields for solving the gyro center's equations of motion. Homogeneous Dirichlet boundary conditions are applied by setting the boundary values to zero. 
The matrices and the terms on the right-hand side are as follows,
\begin{eqnarray}
    \bar{\bar{M}}_{\mathrm{P},L,ii',jj',kk'}
    &&=-\sum n_{0s} \bar{m}_s \frac{B^2_{\rm{ref}}}{B^2}\int {\rm d} R\, {\rm d}Z\, {\rm d}\phi\, J  \nabla_\perp\Upsilon_{ijk} 
    \boldsymbol{\cdot}\nabla_\perp \Upsilon_{i'j'k'}\;\;,
    \nonumber \\ 
    \delta N^{i,j}
    &&=
    -\sum \Bar{q}_s\int {\rm d} R\, {\rm d} Z\, {\rm d}\phi\, J \delta\bar{N}_s(R,\phi,Z) \Upsilon_{ijk} \;\;,\nonumber
\end{eqnarray}
\begin{eqnarray}
    \bar{\bar{M}}_{\mathrm{A},L,ii',jj',kk'}
    &&=\int {\rm d} R\, {\rm d}Z\, {\rm d}\phi\, J
    \left(
    -\nabla_\perp\Upsilon_{ijk} 
    \boldsymbol{\cdot}\nabla_\perp \Upsilon_{i'j'k'}
    -\sum_s\frac{\bar{q}_s^2}{\bar m_s}C_{\mathrm A} 
    \Upsilon_{ijk}\Upsilon_{i'j'k'} \right) \nonumber\;\;,
\end{eqnarray}
\begin{eqnarray}
\label{eq:matrix_form}
    \bar{\bar{M}}_{\mathrm{A},R,ii',jj',kk'}
    &&=\int {\rm d} R\, {\rm d}Z\, {\rm d}\phi\, J
    \nabla_\perp\Upsilon_{ijk} 
    \boldsymbol{\cdot}\nabla_\perp \Upsilon_{i'j'k'}\;\;,
    \nonumber\\
    \delta J^{i,j}_\|
    &&= 
    - \int {\rm d} R\, {\rm d} Z\, {\rm d}\phi\, J \delta\bar J_\|(R,\phi,Z) \Upsilon_{ijk} 
 \;\;,\nonumber\\
    \bar{\bar{M}}_{\mathrm{it},L,ii',jj',kk'}
    &&=
    \bar{\bar{M}}_{\mathrm{A},L,ii',jj',kk'} \;\;, \nonumber \\
    \overline{\langle\delta A^{\rm h}_{i,j,I} \rangle}  
    &&=
    \int {\rm d}R\,{\rm d}Z\,{\rm d}\phi\, J\overline{\langle\delta A^{\rm h}_{I} \rangle} \Upsilon_{ijk}
    \;\;,\\
    \bar{\bar{M}}_{\mathrm{it},R,ii',jj',kk'}
    &&=
    -C_{\mathrm A} \sum_s\frac{\bar{q}_s^2}{\bar m_s}\int {\rm d} R\, {\rm d} Z\, {\rm d}\phi\, J 
    \Upsilon_{ijk}\Upsilon_{i'j'k'} \;\;,\nonumber \\
     \bar{\bar{M}}_{{\rm Ohm},L,ii',jj',kk'} 
     &&=
    \int {\rm d} R\, {\rm d} Z\, {\rm d}\phi\, J 
    \Upsilon_{ijk}\Upsilon_{i'j'k'}\;\;,
    \nonumber \\
     \bar{\bar{M}}_{{\rm Ohm},R,ii',jj',kk'} 
     &&=
    -\int {\rm d} R\, {\rm d} Z\, {\rm d}\phi\, J 
    \Upsilon_{ijk} \partial_\|\Upsilon_{i'j'k'} \;\;,
    \nonumber 
\end{eqnarray}
where $\Upsilon_{ijk}$ is chosen as the traditional 3D basis function or the piecewise field-aligned basis function as discussed in Section~\ref{subsec:solvers}. }

\section{Simulation results}
\label{sec:results}
\subsection{Parameters and setup of the TCV-X21 case}
\label{subsec:parameters}
The TCV-X21 case has been studied  experimentally and numerically for Tokamak à configuration variable (TCV) \cite{oliveira2022validation,body2022development,ulbl2023influence}. This case is used by various codes such as GRILLIX and GENE-X for the studies of the transport and the profile generation with the consideration of the separatrix. The simulation starts from the EQDSK file that gives the poloidal flux function $\psi_{\mathrm p}$, and other profiles and parameters such as the safety factor $q(\psi_{\mathrm p})$. The upper left frame of Fig. \ref{fig:eqdsk_mesh} shows the poloidal flux function $\psi_{\mathrm p}$ where the solid and dashed red lines indicate the vessel surface and the plasma separatrix, respectively. The profile of the safety factor is shown in the lower left frame. The simulation domain is chosen as shown in the upper right frame. The triangular mesh is plotted with reduced resolution (1476 vertices) compared to that used in the simulations (5037 vertices) for better visibility. 
The vertices of the triangular mesh are aligned with the flux surfaces, with additional refinement applied in the open field line region, especially in the vicinity of the X-point, to accurately capture the complex geometry.

%%%%%%%%%%%%%%%%%% psi and mesh%%%%%%%%%%%%%%%%%%%
\begin{figure*}[t]
\centering
\includegraphics[width=0.4\textwidth]{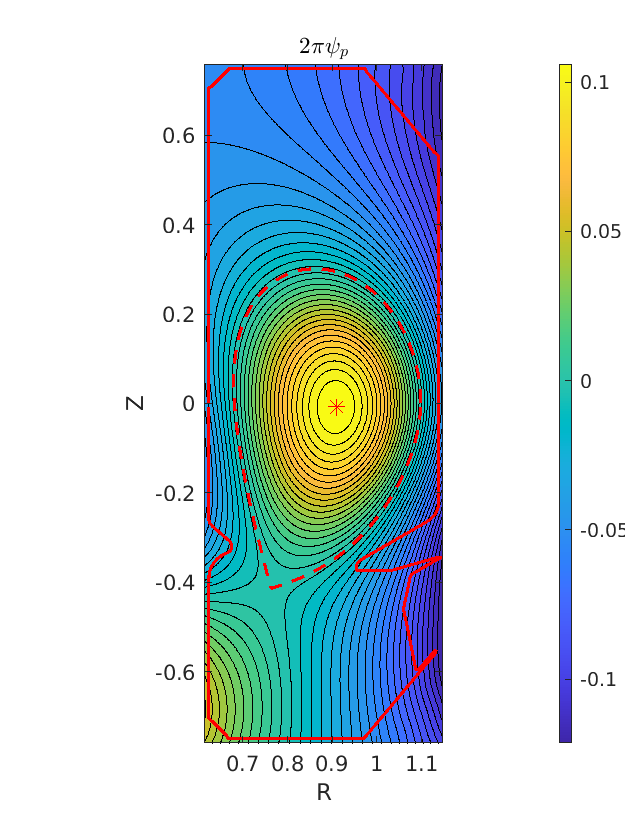}
\includegraphics[width=0.4\textwidth]{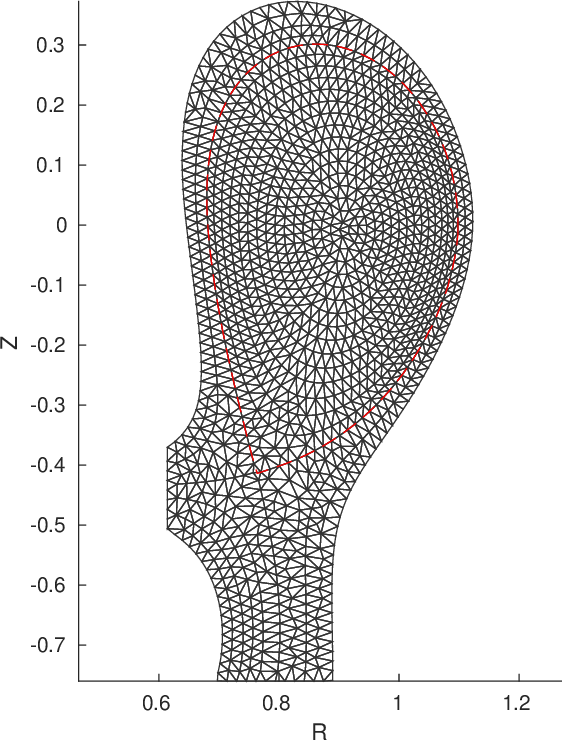}
\includegraphics[width=0.8\textwidth]{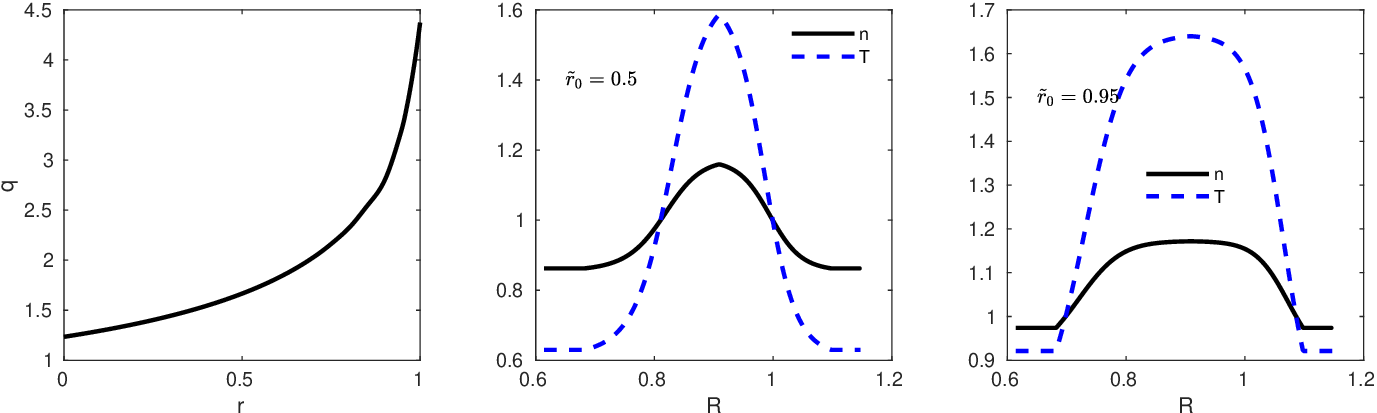}
\caption{The poloidal flux function \(2\pi\psi_{\mathrm p}\) (upper left); the simulation domain with triangular meshes (upper right); the safety factor profile (lower left); the density and temperature (\(n, T\)) profiles along $R$ at $Z=Z_{\rm axis}$ for simulations of the core instability (lower middle) and the edge instability (lower right).
The density and temperature are normalized to $n_{\rm ref}$ and $T_{\rm ref}$, respectively (Section \ref{subsec:parameters}).}
\label{fig:eqdsk_mesh}
\end{figure*}

The simulation is performed in this section to demonstrate the basic features of the linear ITG instability in the presence of the open field lines. For single-$n$ simulations, the particle-in-Fourier scheme is adopted in calculating the density and the current perturbation, but converted to the finite element space before the field equations are solved \cite{lu2026cpc}. 
The reference plasma beta $\beta_{\rm ref}={\mu_0 n_{\rm ref} m_{\rm N} v_{\rm N}^2}/{B_{\rm ref}^2}$ and the reference Larmor radius ${\rho_{\rm ref} }={m_{\rm N}v_{\rm N}}/{(eB_{\rm ref})}$, where $m_N$ is the proton mass, $v_N=\sqrt{2T_{\rm ref}/m_N}$, and the subscripts ``$N$'' and ``ref'' denote the normalization unit and some reference parameter-value in the TRIMEG-C1 code. The reference magnetic field is $B_{\rm{ref}}=1\; \rm{T}$ while the on-axis magnetic field is $B_{\rm axis}=0.90727$ T. The reference density and temperature are $n_{\rm ref}=3.5\times 10^{19}/m^3$ and $T_{\rm ref}=600$ eV, respectively, at the reference radial location $\tilde r_0=0$.  Correspondingly, the reference Larmor radius is $\rho_{\rm ref}=0.3539$ cm and reference beta is $\beta_{\rm ref}=0.84561\%$. More details of the normalization in TRIMEG-C1 are in the previous work \cite{lu2023full,lu2024gyrokinetic}. 
The ion-to-electron mass ratio used is 100. The equilibrium density and temperature profiles ${H(\tilde r)}$, with $H=n$ or $T$, are given by 
\begin{eqnarray}
\label{eq:nTprofile}
     \frac{H(\tilde r)}{H(\tilde r_0)}  =\exp\left[-\kappa_\mathrm{H}w_\mathrm{H}\frac{\tilde{a}}{R_0}\tanh\left(\frac{\tilde r-\tilde r_0}{w_\mathrm{H}}\right)\right]\;\;,  
\end{eqnarray}
where $\tilde{r}=\sqrt{(\psi-\psi_{\rm axis})/(\psi_{\rm edge}-\psi_{\rm axis})}$, $\kappa_{T}=6.96$, $\kappa_{n}=2.23$, the center of the simulation domain $R_0=0.88$ m, $\tilde{a}=(R_{\rm sep, max}-R_{\rm sep, min}))/2=0.20926$ m, $R_{\rm sep}$ is the $R$ coordinate of the separatrix, and $w_{n,T}=0.3$. To study the instabilities destabilized in the core plasma and near the separatrix, we selected $\tilde{r}_0=0.5$ and $ 0.95$, respectively. The density and temperature are assumed to be constant outside of the separatrix in this work for the sake of simplicity. The equilibrium ion and electron profiles are assumed to be identical. The density and the temperature profiles are shown in the lower frame of Fig.~\ref{fig:energy_multi_n}. The finite Larmor radius effect is omitted in this work. To study features of the instabilities with different normalized machine size $a/\rho_{\rm ref}$, another case with $\rho_{\rm ref}=0.01$ m is chosen, which corresponds to $n_{\rm ref}=0.43847\times 10^{19}/m^3$ and $T_{\rm ref}=4789.4$ eV. Four cases studied in the following subsections are summarized in Tab. \ref{tab:cases}. In the following, Sections \ref{subsec:tcv_em_edge}--\ref{subsec:tcv_rho1cm} concentrate on the analysis of linear properties, while Section~\ref{subsec:tcv_multi_n_edge} addresses the non-linear dynamics. 
\begin{table}[h!]
\centering
\begin{tabular}{c c c c c}
\hline\hline
  & $\rho_{\rm ref}$ (cm) &  $\tilde{r}_0$ \\
\hline 
1a & $0.3539$ & 0.5 \\ \hline 
1b & $0.3539$ & 0.95\\ \hline 
2a & $1.0$    & 0.5\\ \hline 
2b & $1.0$    &0.95\\
\hline 
\end{tabular}
\caption{\label{tab:cases} Four cases with different values of $\rho_{\rm ref}$ and $\tilde{r}_0$ are used for the numerical studies of the TCV-X21 scenario.}
\end{table}
%=============================================================
%\input{result_electrostatic}

%=============================================================
\subsection{Electromagnetic instabilities in the core and near separatrix}
\label{subsec:tcv_em_edge}
%The electromagnetic simulation is performed with the parameters $\rho_{\rm ref}=0.35394$ cm, $\beta_{\rm ref}= 0.84561\%$, which corresponds to $n_{\rm ref}=3.5\times10^{19}/{\rm m}^3$, $T_{\rm ref}=600 {\rm eV}$. 
To demonstrate the performance of TRIMEG-C1 for electromagnetic simulations, the instabilities in the core and near the separatrix are simulated using the parameters 1a and 1b in Tab. \ref{tab:cases}. 
The 2D mode structures of $\delta\phi$ and $\delta A_\|$ are displayed in Fig.~\ref{fig:mode2d_em_tcv} for the $n=10$ mode with $\rho_{\rm ref}=0.3539$ cm. The time step size is $\Delta t/t_N=0.001$ where $t_N$ is the time unit of the TRIMEG-C1 code \cite{lu2024gyrokinetic,lu2026cpc}. The results for gradient-driven instabilities in the core are shown in the upper panel of Fig.~\ref{fig:mode2d_em_tcv} (case 1a), while those for the edge region near the separatrix are presented in the lower panel (case 1b).
In case 1a, where the density and temperature gradients are located at $\tilde{r}_0 = 0.5$, the electrostatic potential fluctuation $\delta\phi$ exhibits a clear ballooning structure, while the magnitude of the parallel magnetic potential $\delta A$ is lower at the low field side, due to the different property between $\delta A$ and $\delta \phi$ along the magnetic field line. The perturbation $\delta\phi$ typically peaks near $\theta=0$ and exhibits even parity with respect to $\theta = 0$ along the magnetic field line, leading to localization at the outboard midplane.  However,  for ITG mode, $\delta A_\|$ tends to have an odd parity with respect to $\theta = 0$ along the magnetic field line, resulting in a reduced amplitude at the low-field-side midplane, which is common in Alfv\`enic or electrostatic micro-scale  instabilities \cite{chen2016physics,zonca2014theory,lu2012theoretical}.

In case 1b, where the gradients are shifted outward to $\tilde{r}_0 = 0.95$ near the separatrix, both $\delta\phi$ and $\delta A_\|$ are localized closer to the plasma separatrix and become more radially extended. Compared to the core case, the modes near the separatrix display poloidal variation and finer-scale structure, reflecting the enhanced magnetic shear effects in the edge region. The relative magnitude of $\delta A_\|$ compared with $\delta \phi$ remains comparable between cases 1a and 1b, indicating that the relative importance of electromagnetic effects is similar in the two cases.

%\guo{Can you add a figure showing the q profile and n T profile??? Figure 3 2D mode structure, Please check the explanation the upper and lower. }

The growth rate and the ratio of the electromagnetic and electrostatic field energy are calculated as shown in Fig. \ref{fig:compare_growth_edge_core_nscan}. In the left frame, the growth rate increases as $n$ increases and reaches a saturation level, and then decreases. In the right frame, the field energy ratio $E_A/E_\phi$ quantifies the relative contribution of the electromagnetic field energy associated with $\delta A_\parallel$, where $E_A$ and $E_\phi$ are the  field energy associated with $\delta A_\parallel$ and $\delta\phi$ as defined in the previous work \cite{lu2024gyrokinetic}. For low $n$ modes, the electromagnetic effect is more significant as shown by the higher value of $E_A/E_\phi$. 
In both the growth rate and the energy ratio, the results of the instabilities in the core and near the separatrix are similar since the $n$ and $T$ profiles are similar near $\tilde{r}_0$ and the difference are mainly due to the contribution from other equilibrium parameters such as the local safety factor, the magnetic field strength at different minor-radius locations.

%%%%%%%%%%%%%%%%%% 2D Linear mode structures%%%%%%%%%%%%%%%%%%%
\begin{figure*}[t]
\centering
\includegraphics[width=0.7\textwidth]{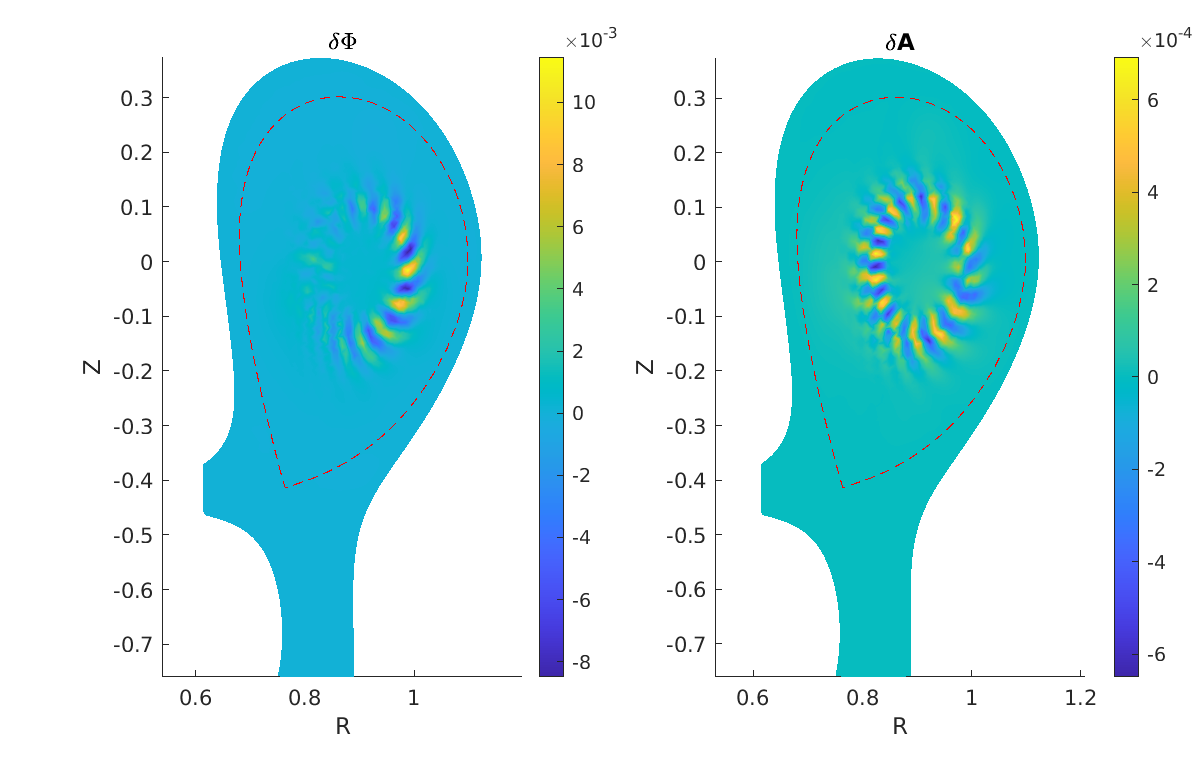}
\includegraphics[width=0.7\textwidth]{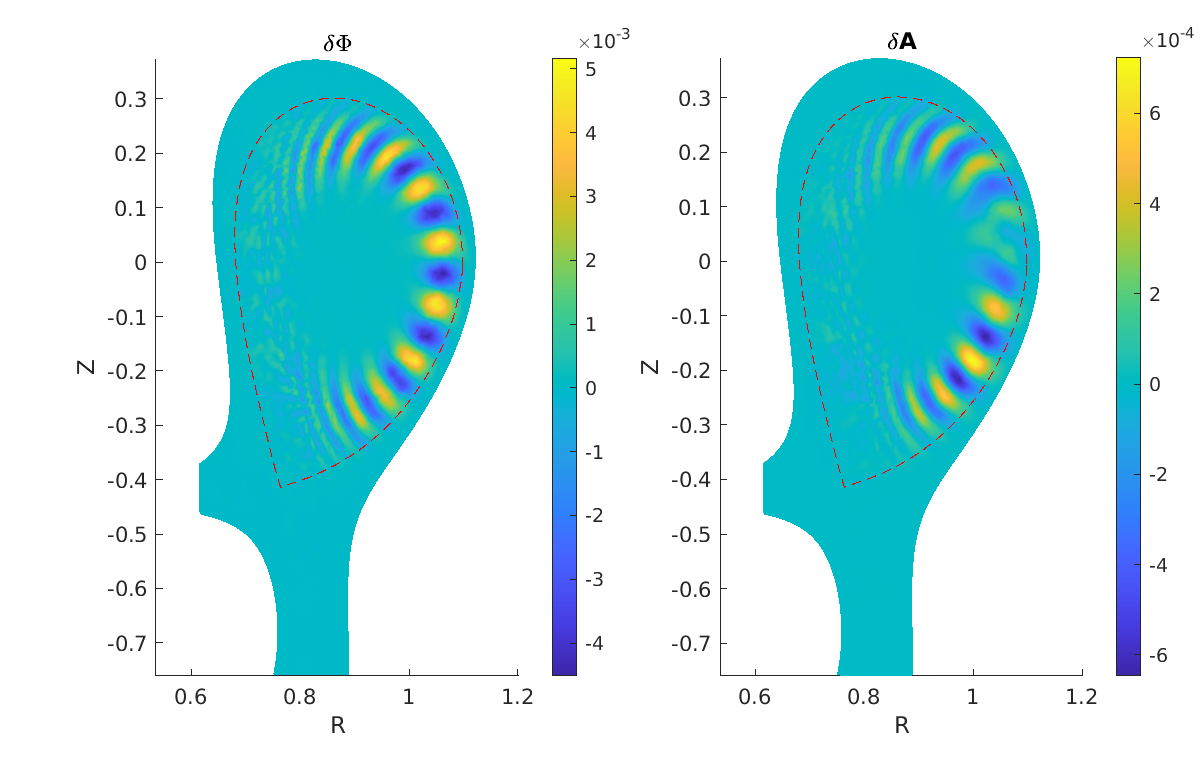}
\caption{The 2D structures of $\delta\phi$ and $\delta A_\|$ for the $n=10$ ITG mode  (Section \ref{subsec:tcv_em_edge}). Top row: case 1a; bottom row: case 1b. }
\label{fig:mode2d_em_tcv}
\end{figure*}

%%%%%%%%%%%%%%%%%% n scan, core&edge %%%%%%%%%%%%%%%%%%%
\begin{figure*}[t]
\centering
\includegraphics[width=0.9\textwidth]{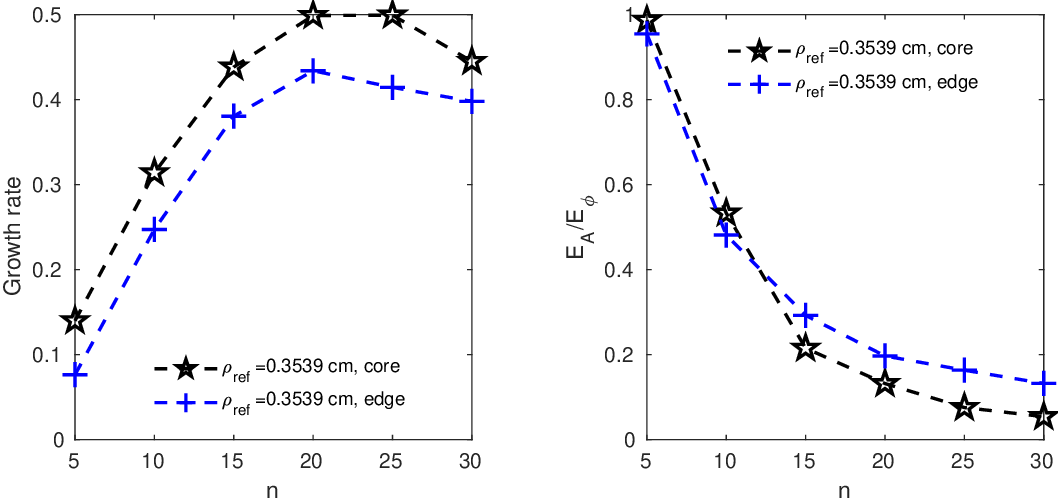}
\caption{ The growth rate (left) and the ratio of the field energy (right) across the spectrum of toroidal mode number $n$ for instabilities in the core (case 1a) and near the separatrix (case 1b) (Section \ref{subsec:tcv_em_edge}).}
\label{fig:compare_growth_edge_core_nscan}
\end{figure*}

%=============================================================
\subsection{Electromagnetic simulations for various $\beta$}
\label{subsec:tcv_em_beta}
%%%%%%%%%%%%%%%%%% beta scan, core%%%%%%%%%%%%%%%%%%%
\begin{figure*}[t]
\centering
\includegraphics[width=0.9\textwidth]{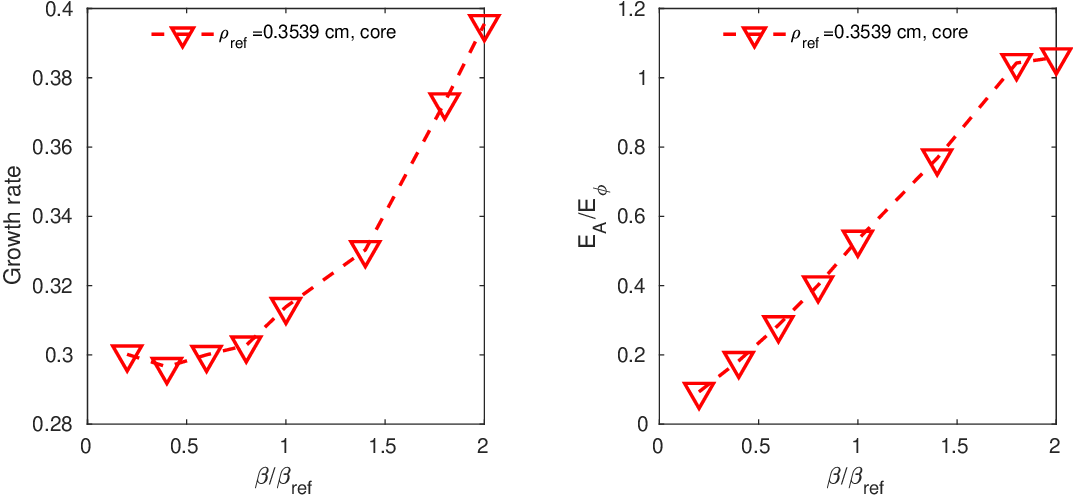}
\caption{ The growth rate (left) and the ratio of the field energy (right) for different values of $\beta_{\rm ref}$ ($\beta$ scan based on case 1a) for the $n=10$ harmonic. (Section \ref{subsec:tcv_em_beta}).}
\label{fig:compare_growth_core_betascan}
\end{figure*}

The relevance of the electromagnetic effects is evaluated by varying the value of the plasma $\beta$. As $\beta$ increases, the dominant instability can change from the ITG mode to the KBM according to previous studies \cite{zonca1999existence,aleynikova2017quantitative,gorler2016intercode}. The $\beta$ effect on the $n=10$ harmonic of case 1a is shown in Fig. \ref{fig:compare_growth_core_betascan}. It suggests that in the ITG mode transits to a KBM as $\beta/\beta_{\rm ref}$ changes from $0.2$ to $2.0$. The nominal $\beta=\beta_{\rm ref}$ is close to the ITG-KBM transition threshold for $n=10$. It should be noted that the KBM-ITG transition relies on multiple parameters such as $\beta$ and $n$, thus for different parameters such as $n$, the critical $\beta$ can be different. In the right frame, the field energy ratio increases with $\beta$, indicating stronger electromagnetic effects. For the nominal value of $\beta$, the field energy ratio shows that the electromagnetic field energy associated with $\delta A_\parallel$ is significant, and the transport due to the electromagnetic effect should be verified. 

%=============================================================
\subsection{Instability with different normalized machine size ($a/\rho_{\rm ref}$)}
\label{subsec:tcv_rho1cm}
Simulations with smaller normalized machine size are carried out using cases 2a and 2b in Tab. \ref{tab:cases}. The growth rate and the energy ratio are shown in Fig. \ref{fig:compare_growth_edge_nscan_rho1cm}. Compared with the results for $\rho_{\rm ref}=0.3539$ cm (cases 1a and 1b) shown in Fig. \ref{fig:compare_growth_edge_core_nscan}, the $\gamma-n$ spectrum downshifts and thus the instability is dominated by modes with lower $n$.
The energy ratio $E_A/E_\phi$ is smaller compared with Fig. \ref{fig:compare_growth_edge_core_nscan}. As $\rho_{\rm ref}$ increases, the energy ratio decreases, indicating a weaker electromagnetic effect, which agrees with the observation in the 1D model (Fig. 1 of \cite{lu2025generalized}).

For numerical studies, higher $\rho_{\rm ref}$ brings in benefit since as $\rho_{\rm ref}/a$ increases, the marker number can be smaller and the time step size can be larger to achieve the same simulation quality, as observed previously in TRIMEG-C1 \cite{lu2024gyrokinetic}. In the multi-$n$ nonlinear simulations in Section \ref{subsec:tcv_multi_n_edge}, we will use  $\rho_{\rm ref}=1$ cm.

%%%%%%%%%%%%%%%%%% n scan, edge, rho=1cm %%%%%%%%%%%%%%%%%%%
\begin{figure*}[htbp]
\centering
\includegraphics[width=0.9\textwidth]{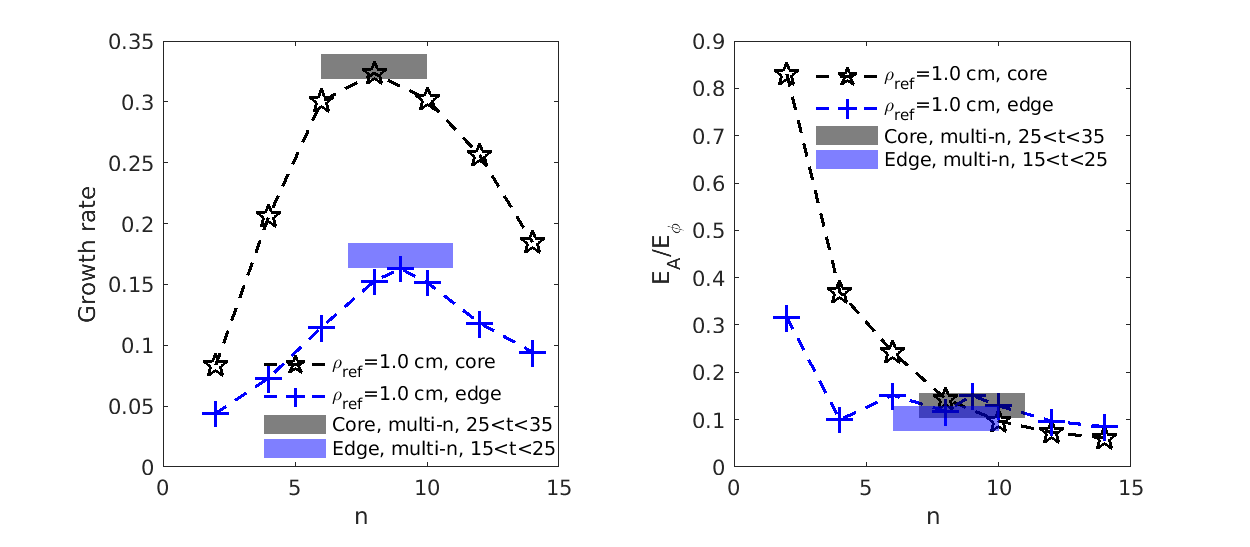}
\caption{ The growth rate (left) and the ratio of the field energy (right) across the toroidal mode spectrum with $\rho_{\rm ref}=1$ cm for core (case 2a) and edge (case 2b) instabilities (Section \ref{subsec:tcv_rho1cm}). The growth rate and the ratio of the field energy during the linear stage in multi-$n$ simulations are indicated by the wide black and blue transparent bars (Section \ref{subsec:tcv_multi_n_edge}).  }
\label{fig:compare_growth_edge_nscan_rho1cm}
\end{figure*}

%=============================================================
\subsection{Multi-$n$ nonlinear simulations}
\label{subsec:tcv_multi_n_edge}
%/tokp/work/luzhixin/2022/gkx-c1/tcv20250914edge1fa1/nallne008e6me.02ngi1bt.0008rho.010nr32c1em1nphi16ohm0fixvparitgT95
Multi-$n$ nonlinear simulations of case 2a and case 2b are performed using a piecewise field-aligned solver without toroidal filtering. The simulations cover the full toroidal domain, $\phi \in [0, 2\pi)$.  In the simulation of the core/edge instability, $\{8,8\}\times10^6$ electron markers and $\{4,2\}\times10^6$ ion markers are simulated. The time step size was chosen as  $\Delta t/t_N=\{0.005,0.004\}$. Completing $\sim 20000$ steps on 4 nodes (1024 cores) of a cluster with AMD EPYC 9754 128-Core Processors (2 processors per node) took $\sim\{72,60\}$ hours. The time evolution of the total field energy and the field energy ratio is demonstrated in Fig. \ref{fig:energy_multi_n}. The instability growth rates $\gamma_{AP}$, $\gamma_A$, $\gamma_P$ are calculated from the total field energy $E_{\rm tot}=E_\phi+E_A$,  $E_A$ and $E_\phi$, respectively. These energies are assumed to evolve as $E_{[\rm{tot},A,\phi]}=E_{[\rm{tot},A,\phi]}(t=0) \exp\{2\gamma_{[AP,A,P]} t\}$ at linear stage. The growth rates are measured during the linear stage $25<t/t_N<35$ for case 2a and $15<t/t_N<25$ for case 2b. As shown in the right frame of Fig. \ref{fig:energy_multi_n}, the instability is dominated by the electrostatic component since the electromagnetic energy ratio is relatively low with $E_A/E_{\rm tot}<0.15$. 

The most unstable mode appears and becomes dominant in the late linear stage ($t/t_N\in [25,35]$ for case 2a and $t/t_N\in[15,25]$ for case 2b), and the 2D mode structure is shown in Fig. \ref{fig:mode2d_tcv_nonlinear_multin}. 
The growth rate and the field energy ratio measured during the linear stage are also plotted in Fig.~\ref{fig:compare_growth_edge_nscan_rho1cm} for comparison with the single-$n$ simulation, which are indicated by the blue and black transparent bars. The dominant mode numbers $n\sim 6-10$ for case 2a and $n \sim 7-11$ for case 2b in the multi-$n$ simulations are indicated by the widths of the bars. The growth rate and the energy ratio are similar to those of the most unstable mode from the single-$n$ simulations, while the minor difference is due to the finite error of the measurement. 
%%%%%%%%%%%%%%%%%% time evolution of total energy, rho=1cm %%%%%%%%%%%%%%%%%%%
\begin{figure*}[t]
\centering
\includegraphics[width=0.9\textwidth]{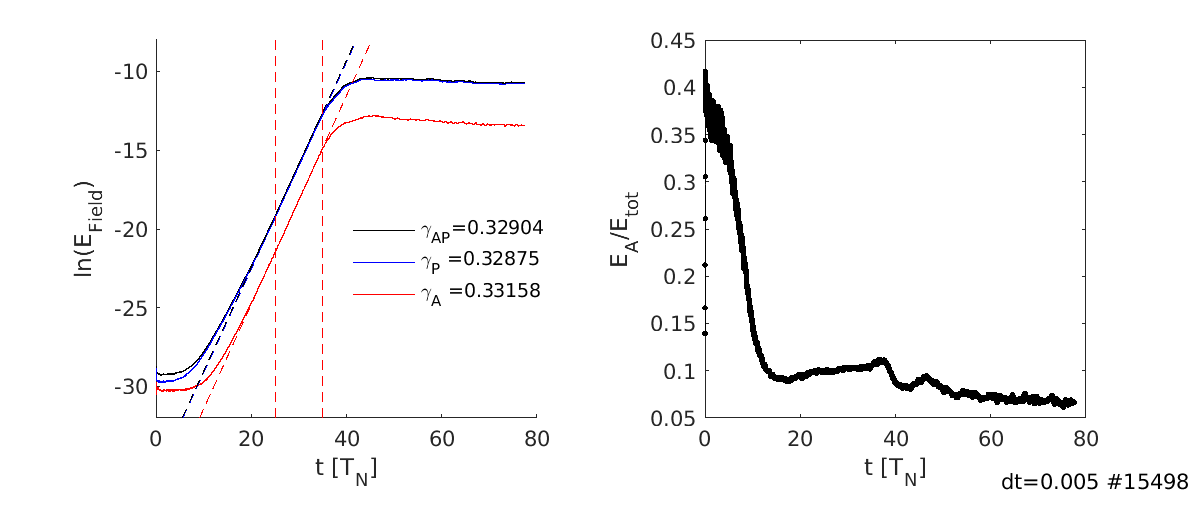}
\includegraphics[width=0.9\textwidth]{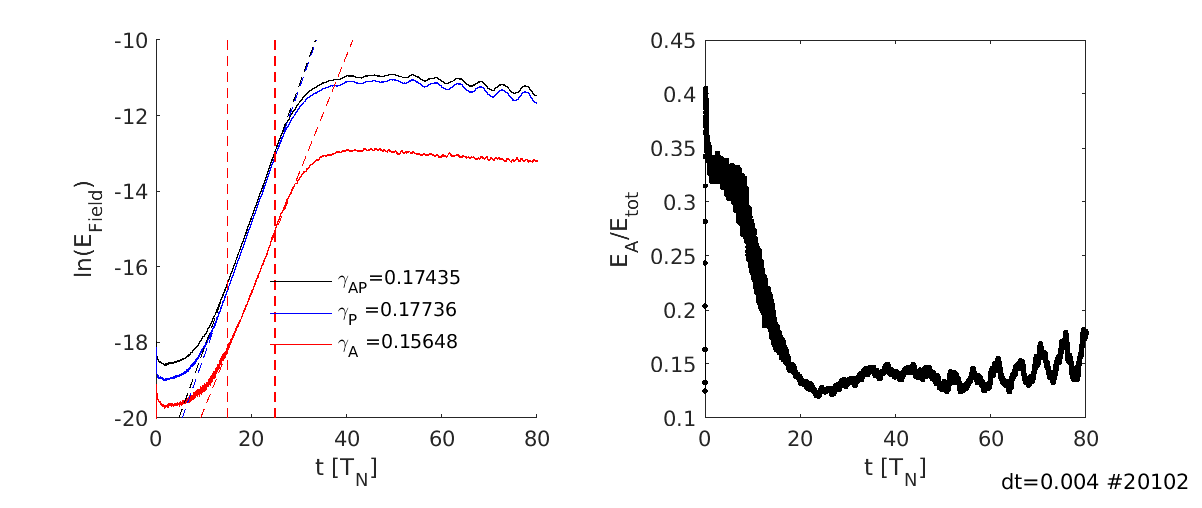}
\caption{ The time evolution of the total field energy (left) and the field energy ratio (right) for the core instability (upper frame) and the edge instability (lower frame) in the nonlinear multi-harmonic simulations (Section \ref{subsec:tcv_multi_n_edge}).}
\label{fig:energy_multi_n}
\end{figure*}

%%%%%%%%%%%%%%%%%% time evolution of total energy, rho=1cm %%%%%%%%%%%%%%%%%%%
\begin{figure*}[t]
\centering
\includegraphics[width=0.7\textwidth]{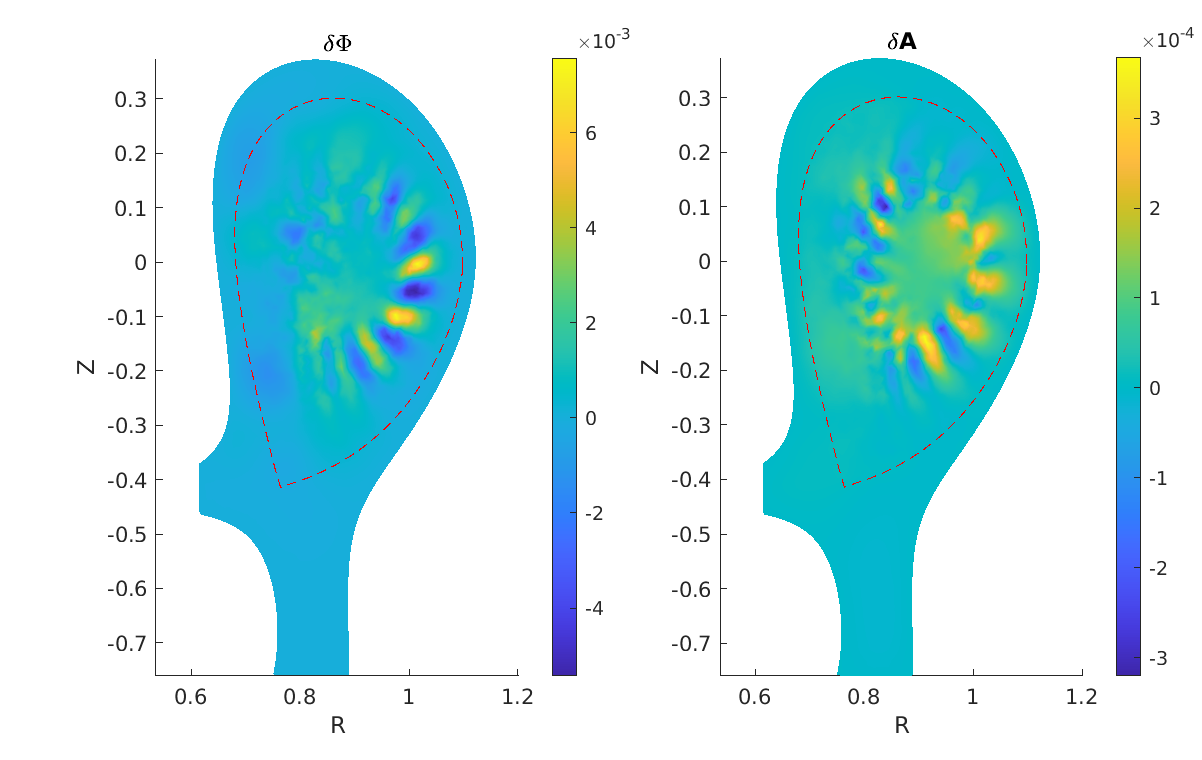}
\includegraphics[width=0.7\textwidth]{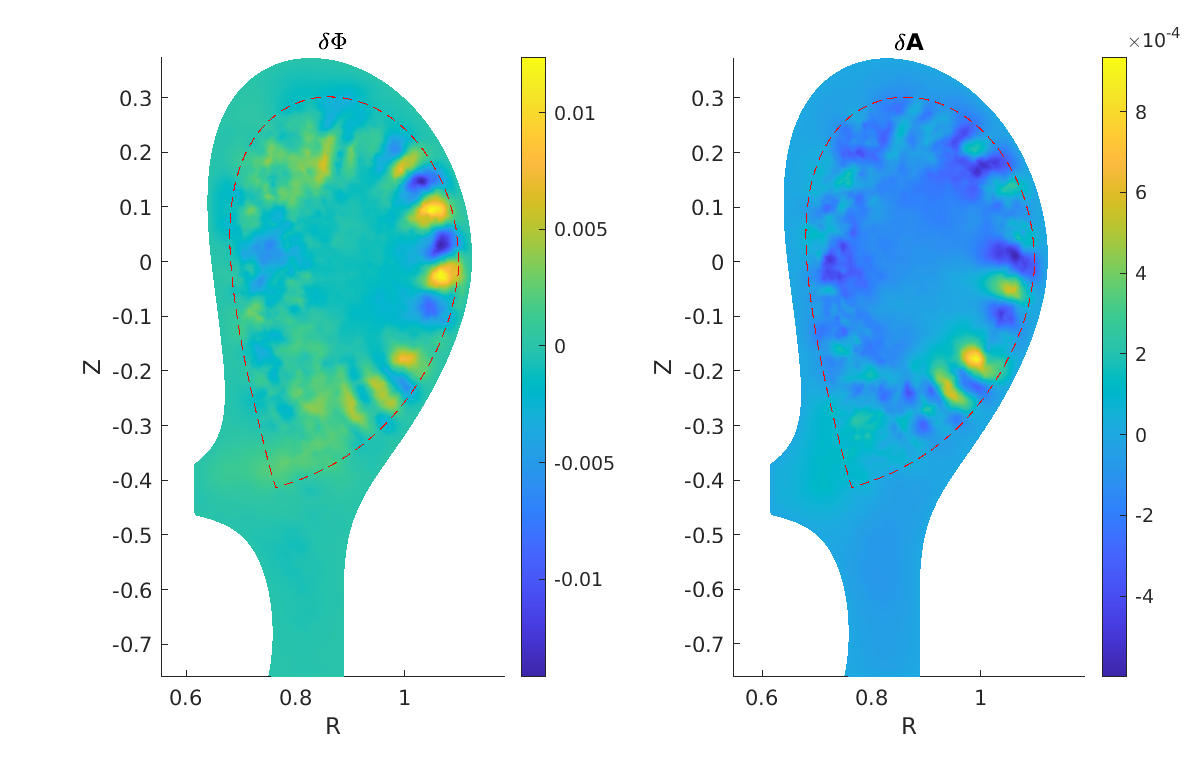}
\caption{ The 2D mode structures of the core instability ($t/t_N=35$, upper frame) and the edge instability ($t/t_N=30$, lower frame) at the late linear stage of multi-$n$ nonlinear simulations (Section \ref{subsec:tcv_multi_n_edge}).}
\label{fig:mode2d_tcv_nonlinear_multin}
\end{figure*}

%******************************************************************************
\section{Conclusions}
\label{sec:conclusion}
In this work, a high-order  $C^1$-continuous piecewise field-aligned finite element method on unstructured triangular meshes has been extended and implemented in the TRIMEG-C1 gyrokinetic particle-in-cell framework for whole-volume electromagnetic simulations of tokamak plasmas with open field lines. By combining $C^1$ triangular finite elements, mixed-variable formulations, and a generalized pullback scheme, the developed approach enables efficient and stable simulations of electromagnetic micro instabilities in complex geometries spanning the core, separatrix, and open field line regions.

The piecewise field-aligned finite element method formulated in cylindrical coordinates significantly reduces grid distortion associated with magnetic shear while retaining field-line alignment locally within each toroidal subdomain. This feature proves essential for whole-volume simulations, where global field-aligned coordinates become impractical. The formulation is fully compatible with both $\delta f$ and full-f models and supports the iterative solution of Amp\`ere's law with analytically well-controlled skin-depth terms, ensuring robust convergence in electromagnetic regimes.

Linear simulations of the TCV-X21 configuration demonstrate that the developed scheme accurately captures electromagnetic ion-temperature-gradient and kinetic ballooning mode physics. Systematic parameter scans in toroidal mode number, plasma $\beta$, and normalized machine size confirm expected physical trends, including the ITG–KBM transition and the downshift of the dominant instability spectrum at reduced $a/\rho_{\rm ref}$. Multi-$n$ nonlinear simulations further validate the capability of the piecewise field-aligned solver to handle fully three-dimensional electromagnetic turbulence without toroidal filtering.

Overall, the results establish TRIMEG-C1 with the present high-order piecewise field-aligned finite element method as a powerful and flexible tool for global electromagnetic gyrokinetic simulations in realistic tokamak geometries with open field lines. Future work will focus on incorporating finite Larmor radius effects, extending the approach to fully nonlinear electromagnetic turbulence in the edge and scrape-off layer, and performing quantitative validation against experimental measurements.

\ack{
The discussion with A. Stegmeir and Ph. Ulbl on the TCV case and the discussion with M. Borchardt are appreciated.
The simulations in this work were run on the local TOK cluster and the MPCDF Viper/Raven supercomputers. The EUROfusion projects TSVV-8, ACH/MPG, TSVV-10, TSVV-G, and ATEP are acknowledged. 
This work has been carried out within the framework of the EUROfusion Consortium, funded by the European Union via the Euratom Research and Training Programme (Grant Agreement No 101052200—EUROfusion). Views and opinions expressed are however those of the author(s) only and do not necessarily reflect those of the European Union or the European Commission. Neither the European Union nor the European Commission can be held responsible for them.

}

%\clearpage
%\null

%\appendix
%\section{}
%\label{}

%\funding{Sample text inserted for demonstration.}
% This section is a list of funder names and grant numbers

%\roles{Sample text inserted for demonstration.}
% List author names and the contributions made to the article, using terms from the NISO Contributor Roles Taxonomy (CRediT) https://credit.niso.org

%\data{Sample text inserted for demonstration.}
% For more information on IOP Publishing's research data policy see: https://publishingsupport.iopscience.iop.org/questions/research-data/

%\suppdata{Sample text inserted for demonstration.}

%\section*{References}

%\bibliographystyle{elsarticle-num-names} 
\bibliographystyle{iopart-num}
\bibliography{reference}

\end{document}